\documentclass[conference]{IEEEtran}
\usepackage[top=0.705in, bottom=1.09in, left=0.65in, right=0.65in]{geometry}
\usepackage{url}
\usepackage[utf8]{inputenc}
\usepackage{xcolor}
\usepackage{amsmath}

\usepackage{multirow}
\usepackage{amssymb}
\usepackage{cite}
\usepackage{enumitem}
\usepackage{bm}
\usepackage{graphicx}
\usepackage{amsthm}

\usepackage{gensymb}

\usepackage[acronyms,nonumberlist,nopostdot,nomain,nogroupskip]{glossaries}
\usepackage{tablefootnote}
\usepackage{booktabs}
\usepackage{tabularx}
\usepackage{epsfig}
\usepackage{tikz}
\usetikzlibrary{external}
\usepackage{pgfplots}
\pgfplotsset{compat=newest} 
\pgfplotsset{plot coordinates/math parser=false}

\usetikzlibrary{plotmarks,patterns,patterns.meta,decorations.pathreplacing,backgrounds,calc,arrows,arrows.meta,spy,matrix}
\usepgfplotslibrary{patchplots,groupplots}
\usepackage{tikzscale}
\usepackage{siunitx}
\usepackage{tablefootnote}

\usepackage{multirow}
\usepackage{algorithm2e, setspace}
\SetAlCapNameFnt{\footnotesize}
\SetAlCapFnt{\footnotesize}
\RestyleAlgo{ruled}
\SetKwComment{Comment}{/* }{ */}
\usepackage{algpseudocode}

\usepackage[font=scriptsize]{subcaption}
\usepackage[font=footnotesize]{caption}

\usepackage{mathtools}

\usepackage{dblfloatfix}    
\usepackage{colortbl}
\usepackage{flushend}
\usepackage{footnote}

\usepackage{lipsum,afterpage,refcount}

\newacronym{3gpp}{3GPP}{3rd Generation Partnership Project}
\newacronym{adc}{ADC}{Analog to Digital Converter}
\newacronym{5g}{5G}{5th generation}
\newacronym{6g}{6G}{6th generation}
\newacronym{ai}{AI}{Artificial Intelligence}
\newacronym{aimd}{AIMD}{Additive Increase Multiplicative Decrease}
\newacronym{am}{AM}{Acknowledged Mode}
\newacronym{amc}{AMC}{Adaptive Modulation and Coding}
\newacronym{aqm}{AQM}{Active Queue Management}
\newacronym{awgn}{AGWN}{Additive White Gaussian Noise}
\newacronym{balia}{BALIA}{Balanced Link Adaptation}
\newacronym{bdp}{BDP}{Bandwidth-Delay Product}
\newacronym{bf}{BF}{beamforming}
\newacronym{cc}{CC}{Congestion Control}
\newacronym{cdf}{CDF}{Cumulative Distribution Function}
\newacronym{cn}{CN}{Core Network}
\newacronym{cqi}{CQI}{Channel Quality Information}
\newacronym{cp}{CP}{Control Plane}
\newacronym{csirs}{CSI-RS}{Channel State Information - Reference Signal}
\newacronym{dc}{DC}{Dual Connectivity}
\newacronym{dr}{DR}{Data Rate}
\newacronym{vsat}{VSAT}{Very Small Aperture Terminal}
\newacronym{rb}{RB}{Resource Block}
\newacronym{dce}{DCE}{Direct Code Execution}
\newacronym{dci}{DCI}{Downlink Control Information}
\newacronym{udp}{UDP}{User Datagram Protocol}
\newacronym{dl}{DL}{downlink}
\newacronym{fcfs}{FCFS}{first-come-first-served}
\newacronym{dmr}{DMR}{Deadline Miss Ratio}
\newacronym{fspl}{FSPL}{free-space path loss}
\newacronym{dmrs}{DMRS}{DeModulation Reference Signal}
\newacronym{e2e}{E2E}{End-to-End}
\newacronym{ppp}{PPP}{Poission Point Process}
\newacronym{aoi}{AoI}{Area of Interest}
\newacronym{cpu}{CPU}{Central Processing Unit}
 \newacronym{gpu}{GPU}{Graphics Processing Unit}
 \newacronym{tpu}{TPU}{Tensor Processing Unit}
\newacronym{si}{SI}{Study Item}
\newacronym{ecn}{ECN}{Explicit Congestion Notification}
\newacronym{edf}{EDF}{Earliest Deadline First}
\newacronym{enb}{eNB}{eNodeB}
\newacronym{epc}{EPC}{Evolved Packet Core}
\newacronym{es}{ES}{Edge Server}
\newacronym{cav}{CAV}{Connected and Autonomous Vehicle}
\newacronym{fdma}{FDMA}{Frequency Division Multiple Access}
\newacronym{fdd}{FDD}{Frequency Division Duplexing}
\newacronym{upa}{UPA}{Uniform Planar Array}
\newacronym{car}{CAR}{Circular Aperture Reflector }
\newacronym[firstplural=Radio Access Technologies (RATs)]{rat}{RAT}{Radio Access Technology}
\newacronym[firstplural=Radio Access Technology (RTs)]{rt}{RT}{Radio Technology}
\newacronym{fs}{FS}{Fast Switching}
\newacronym{isd}{ISD}{inter-site distance}
\newacronym{ftp}{FTP}{File Transfer Protocol}
\newacronym{gnb}{gNB}{Next Generation Node Base}
\newacronym{harq}{HARQ}{Hybrid Automatic Repeat reQuest}
\newacronym{hetnet}{HetNet}{Heterogeneous Network}
\newacronym{hh}{HH}{Hard Handover}
\newacronym{hol}{HOL}{Head-of-Line}
\newacronym{ia}{IA}{Initial Access}
\newacronym{imt}{IMT}{International Mobile Telecommunication}
\newacronym{iot}{IoT}{Internet of Things}
\newacronym{los}{LOS}{Line of Sight}
\newacronym{lte}{LTE}{Long Term Evolution}
\newacronym{m2m}{M2M}{Machine to Machine}
\newacronym{mac}{MAC}{Medium Access Control}
\newacronym{mc}{MC}{Multi-Connectivity}
\newacronym{mcs}{MCS}{Modulation and Coding Scheme}
\newacronym{mec}{MEC}{Mobile Edge Cloud}
\newacronym{mi}{MI}{Mutual Information}
\newacronym{mimo}{MIMO}{Multiple Input Multiple Output}
\newacronym{mmwave}{mmWave}{millimeter wave}
\newacronym{mptcp}{MPTCP}{Multipath TCP}
\newacronym{mr}{MR}{Maximum Rate}
\newacronym{mss}{MSS}{Maximum Segment Size}
\newacronym{mtd}{MTD}{Machine-Type Device}
\newacronym{mtu}{MTU}{Maximum Transmission Unit}
\newacronym{nfv}{NFV}{Network Function Virtualization}
\newacronym{vnf}{VNF}{Virtualization Network Function}
\newacronym{gv}{GV}{ground vehicle}
\newacronym{gvs}{GVs}{ground vehicles}
\newacronym{vec}{VEC}{Vehicular Edge Computing}
\newacronym{sdn}{SDN}{Software Defined Networking}
\newacronym{nlos}{NLOS}{Non Line of Sight}
\newacronym{nlosb}{NLOSb}{Building Non Line of Sight}
\newacronym{nlosv}{NLOSv}{Vehicle Non Line of Sight}
\newacronym{nr}{NR}{New Radio}
\newacronym{ofdm}{OFDM}{Orthogonal Frequency Division Multiplexing}
\newacronym{pdcch}{PDCCH}{Physical Downlonk Control Channel}
\newacronym{pdcp}{PDCP}{Packet Data Convergence Protocol}
\newacronym{pdsch}{PDSCH}{Physical Downlink Shared Channel}
\newacronym{pdu}{PDU}{Packet Data Unit}
\newacronym{pf}{PF}{Proportional Fair}
\newacronym{pgw}{PGW}{Packet Gateway}
\newacronym{phy}{PHY}{Physical}
\newacronym{pbch}{PBCH}{Physical Broadcast Channel}
\newacronym[plural=\gls{mme}s,firstplural=Mobility Management Entities (MMEs)]{mme}{MME}{Mobility Management Entity}
\newacronym{prb}{PRB}{Physical Resource Block}
\newacronym{pss}{PSS}{Primary Synchronization Signal}
\newacronym{pucch}{PUCCH}{Physical Uplink Control Channel}
\newacronym{pusch}{PUSCH}{Physical Uplink Shared Channel}
\newacronym{rach}{RACH}{Random Access Channel}
\newacronym{ran}{RAN}{Radio Access Network}
\newacronym{red}{RED}{Random Early Detection}
\newacronym{rf}{RF}{Radio Frequency}
\newacronym{rlc}{RLC}{Radio Link Control}
\newacronym{rlf}{RLF}{Radio Link Failure}
\newacronym{rrc}{RRC}{Radio Resource Control}
\newacronym{rrm}{RRM}{Radio Resource Management}
\newacronym{rr}{RR}{Round Robin}
\newacronym{rs}{RS}{Remote Server}
\newacronym{rsrp}{RSRP}{Reference Signal Received Power}
\newacronym{rss}{RSS}{Received Signal Strength}
\newacronym{rtt}{RTT}{Round Trip Time}
\newacronym{rw}{RW}{Receive Window}
\newacronym{rx}{RX}{Receiver}
\newacronym{sa}{SA}{standalone}
\newacronym{sack}{SACK}{Selective Acknowledgment}
\newacronym{sap}{SAP}{Service Access Point}
\newacronym{sch}{SCH}{Secondary Cell Handover}
\newacronym{scoot}{SCOOT}{Split Cycle Offset Optimization Technique}
\newacronym{sdma}{SDMA}{Spatial Division Multiple Access}
\newacronym{sinr}{SINR}{Signal to Interference plus Noise Ratio}
\newacronym{sm}{SM}{Saturation Mode}
\newacronym{snr}{SNR}{Signal-to-Noise Ratio}
\newacronym{son}{SON}{Self-Organizing Network}
\newacronym{ss}{SS}{Synchronization Signal}
\newacronym{srs}{SRS}{Sounding Reference Signal}
\newacronym{sss}{SSS}{Secondary Synchronization Signal}
\newacronym{tb}{TB}{Transport Block}
\newacronym{tcp}{TCP}{Transmission Control Protocol}
\newacronym{tdd}{TDD}{Time Division Duplexing}
\newacronym{tdma}{TDMA}{Time Division Multiple Access}
\newacronym{tfl}{TfL}{Transport for London}
\newacronym{tm}{TM}{Transparent Mode}
\newacronym{prr}{PRR}{Packet Reception Ratio}
\newacronym{trp}{TRP}{Transmitter Receiver Pair}
\newacronym{tti}{TTI}{Transmission Time Interval}
\newacronym{ttt}{TTT}{Time-to-Trigger}
\newacronym{tx}{TX}{Transmitter}
\newacronym{ue}{UE}{User Equipment}
\newacronym{ul}{UL}{uplink}
\newacronym{uml}{UML}{Unified Modeling Language}
\newacronym{um}{UM}{Unacknowledged Mode}
\newacronym{utc}{UTC}{Urban Traffic Control}
\newacronym{vm}{VM}{Virtual Machine}
\newacronym{rsrq}{RSRQ}{Reference Signal Received Quality}
\newacronym{rssi}{RSSI}{Received Signal Strength Indicator}
\newacronym{crs}{CRS}{Cell Reference Signal}
\newacronym{v2v}{V2V}{Vehicle-to-Vehicle}
\newacronym{v2i}{V2I}{Vehicle-to-Infrastructure}
\newacronym{v2n}{V2N}{Vehicle-to-Network}
\newacronym{v2x}{V2X}{Vehicle-to-Everything}
\newacronym{vn}{VN}{Vehicular Node}
\newacronym{dsrc}{DSRC}{Dedicated Short Range Communication}
\newacronym{ci}{CI}{context information}
\newacronym{voi}{VoI}{value of information}
\newacronym{gps}{GPS}{Global Positioning System}
\newacronym{qos}{QoS}{Quality of Service}
\newacronym{qoe}{QoE}{Quality of Experience}
\newacronym{ml}{ML}{Machine Learning}
\newacronym{ahp}{AHP}{Analytic Hierarchy Process}
\newacronym{lidar}{LIDAR}{Light Detection and Ranging}
\newacronym{sumo}{SUMO}{Simulation of Urban MObility}
\newacronym{wave}{WAVE}{Wireless Access in Vehicular Environment}
\newacronym{c-its}{C-ITS}{Connected Intelligent Transportation System}
\newacronym{dash}{DASH}{Dynamic Adaptive Streaming over HTTP}
\newacronym{http}{HTTP}{HyperText Transfer Protocol}
\newacronym{nt}{NT}{Non-Terrestrial}
\newacronym{ntc}{NTC}{non-terrestrial communication}
\newacronym{ntn}{NTN}{Non-Terrestrial Network}
\newacronym{hap}{HAP}{High Altitude Platform}
\newacronym{leo}{LEO}{Low Earth Orbit}
\newacronym{meo}{MEO}{Medium Earth Orbit}
\newacronym{geo}{GEO}{Geostationary Earth Orbit}
\newacronym{uav}{UAV}{Unmanned Aerial Vehicle}
\newacronym{nsat}{nSAT}{Nanosatellite}
\newacronym{ehf}{EHF}{extremely high-frequency}
\newacronym{ioe}{IoE}{Internet of Everyone}
\newacronym{gan}{GaN}{Gallium Nitride}
\newacronym{tle}{TLE}{two-line element}
\newacronym{ecdf}{ECDF}{Empirical Cumulative Distribution Function}
\newacronym{fifo}{FIFO}{First-Input First-Output}
\newacronym{gnss}{GNSS}{Global Navigation Satellite System}
\newacronym{lora}{LoRa}{Long Range}
\newacronym{lpwan}{LPWAN}{Low-Power Wide-Area Network}
\newacronym{ed}{ED}{End Device}
\newacronym{gw}{GW}{Gateway}
\newacronym{ns}{NS}{Network Server}
\newacronym{css}{CSS}{Chirp Spread Spectrum}
\newacronym{ism}{ISM}{Industrial, Scientific, and Medical}
\newacronym{sf}{SF}{Spreading Factor}

\usepackage{tikz}
\usepackage{pgfplots}
\usepackage{tikzscale}
\usepackage{tikz-qtree}

\pgfplotsset{compat=newest}
\pgfplotsset{plot coordinates/math parser=false}
\pgfplotsset{every axis/.append style={
                    label style={font=\scriptsize},
                    tick label style={font=\scriptsize},
                    legend style={font=\scriptsize}
                    }}
\usetikzlibrary{plotmarks,shapes,patterns,decorations.pathreplacing,backgrounds,calc,arrows,arrows.meta,spy,matrix,shadows,trees,positioning,fit}
\usepgfplotslibrary{patchplots,groupplots}

\tikzstyle{startstop} = [rectangle, rounded corners, minimum width=2cm, minimum height=0.5cm,text centered, draw=black]
\tikzstyle{io} = [trapezium, trapezium left angle=70, trapezium right angle=110, minimum width=3cm, minimum height=1cm, text centered, draw=black]
\tikzstyle{process} = [rectangle, minimum width=2cm, minimum height=0.5cm, text centered, draw=black, alignb=center]
\tikzstyle{decision} = [ellipse, minimum width=2cm, minimum height=1cm, text centered, draw=black]
\tikzstyle{arrow} = [thick,<->,>=stealth]
\tikzstyle{line} = [thick,>=stealth]
\tikzstyle{darrow} = [thick,<->,>=stealth,dashed]
\tikzstyle{sarrow} = [thick,->,>=stealth]
\tikzstyle{larrow} = [line width=0.1mm,dashdotted,->,>=stealth]

\pgfkeys{/pgf/number format/.cd,1000 sep={\,}} 

\makeatletter
\def\grd@save@target#1{%
  \def\grd@target{#1}}
\def\grd@save@start#1{%
  \def\grd@start{#1}}
\tikzset{
  grid with coordinates/.style={
    to path={%
      \pgfextra{%
        \edef\grd@@target{(\tikztotarget)}%
        \tikz@scan@one@point\grd@save@target\grd@@target\relax
        \edef\grd@@start{(\tikztostart)}%
        \tikz@scan@one@point\grd@save@start\grd@@start\relax
        \draw[minor help lines] (\tikztostart) grid (\tikztotarget);
        \draw[major help lines] (\tikztostart) grid (\tikztotarget);
        \grd@start
        \pgfmathsetmacro{\grd@xa}{\the\pgf@x/1cm}
        \pgfmathsetmacro{\grd@ya}{\the\pgf@y/1cm}
        \grd@target
        \pgfmathsetmacro{\grd@xb}{\the\pgf@x/1cm}
        \pgfmathsetmacro{\grd@yb}{\the\pgf@y/1cm}
        \pgfmathsetmacro{\grd@xc}{\grd@xa + \pgfkeysvalueof{/tikz/grid with coordinates/major step x}}
        \pgfmathsetmacro{\grd@yc}{\grd@ya + \pgfkeysvalueof{/tikz/grid with coordinates/major step y}}
        \foreach \x in {\grd@xa,\grd@xc,...,\grd@xb}
        \node[anchor=north] at (\x,\grd@ya) {\pgfmathprintnumber{\x}};
        \foreach \y in {\grd@ya,\grd@yc,...,\grd@yb}
        \node[anchor=east] at (\grd@xa,\y) {\pgfmathprintnumber{\y}};
      }
    }
  },
  minor help lines/.style={
    help lines,
    gray,
    line cap =round,
    xstep=\pgfkeysvalueof{/tikz/grid with coordinates/minor step x},
    ystep=\pgfkeysvalueof{/tikz/grid with coordinates/minor step y}
  },
  major help lines/.style={
    help lines,
    line cap =round,
    line width=\pgfkeysvalueof{/tikz/grid with coordinates/major line width},
    xstep=\pgfkeysvalueof{/tikz/grid with coordinates/major step x},
    ystep=\pgfkeysvalueof{/tikz/grid with coordinates/major step y}
  },
  grid with coordinates/.cd,
  minor step x/.initial=.5,
  minor step y/.initial=.2,
  major step x/.initial=1,
  major step y/.initial=1,
  major line width/.initial=1pt,
}
\makeatother

\newlength\fheight
\newlength\fwidth

\definecolor{steelblue}{RGB}{176,196,222}

\usepackage[capitalize]{cleveref}
\crefname{section}{Sec.}{Secs.}
\usetikzlibrary{decorations}

\newcommand\copyrightnotice{%
\begin{tikzpicture}[remember picture,overlay]
\node[anchor=south,yshift=15pt] at (current page.south) {\fbox{\parbox{\dimexpr\textwidth-\fboxsep-\fboxrule\relax}{
\footnotesize \textcopyright 2025 IEEE. Personal use of this material is permitted.
Permission from IEEE must be obtained for all other uses, in any current or future media,
including reprinting/republishing this material for advertising or promotional purposes,
creating new collective works, for resale or redistribution to servers or lists,
or reuse of any copyrighted component of this work in other works.}}};
\end{tikzpicture}
}

\begin{document}
\bstctlcite{IEEEexample:BSTcontrol}

\title{Performance Evaluation of LoRa for IoT Applications in Non-Terrestrial Networks via ns-3}

\author{\IEEEauthorblockN{Alessandro Traspadini, Michele Zorzi, Marco Giordani}
\IEEEauthorblockA{Department of Information Engineering, University of Padova, Italy \\
Email: \texttt{\{traspadini, zorzi, giordani\}@dei.unipd.it}} \vspace{-1cm}
}

\maketitle

\copyrightnotice

\begin{abstract}
The integration of \gls{iot} and \glspl{ntn} has emerged as a key paradigm to provide connectivity for sensors and actuators via satellite gateways in remote areas where terrestrial infrastructure is limited or unavailable.
Among other \gls{lpwan} technologies for IoT, \gls{lora} holds great potential given its long range, energy efficiency, and flexibility.
In this paper, we explore the feasibility and performance of LoRa to support large-scale IoT connectivity through \gls{leo} satellite gateways.
To do so, we developed a new \texttt{ns3-LoRa-NTN} simulation module, which integrates and extends the \texttt{ns3-LoRa} and \texttt{ns3-NTN} modules, to enable full-stack end-to-end simulation of satellite communication in LoRa networks. 
Our results, given in terms of average data rate and \gls{prr}, confirm that LoRa can effectively support direct
communication from the ground to LEO satellites, but network optimization is required to mitigate collision probability when end
nodes use the same \glspl{sf}  over long distances.
\end{abstract}

\glsresetall


\begin{tikzpicture}[remember picture,overlay]
\node[anchor=north,yshift=-10pt] at (current page.north) {\parbox{\dimexpr\textwidth-\fboxsep-\fboxrule\relax}{
\centering\footnotesize 
This paper has been accepted for publication in the 2025 IEEE Global Communications Conference (GLOBECOM) \textcopyright 2025 IEEE.\\
Please cite it as: A. Traspadini, M. Zorzi, and M. Giordani, “Performance Evaluation of LoRa for IoT Applications in Non-Terrestrial Networks via ns-3,” in Proc. IEEE Global Communications Conference (GLOBECOM), 2025.}};
\end{tikzpicture}

\glsresetall

\section{Introduction}
\label{sec:intro}
The global market for the \gls{iot} is expected to reach a value of around USD 11.1 trillion~\cite{Safaei_2017} in 2025, with approximately 75 billion connected units.
To support this growth, current \gls{iot} connectivity solutions primarily rely on \gls{lpwan} technologies such as \gls{lora}~\cite{magrin2017performance}, Narrowband-IoT (NB-IoT)~\cite{MFXAB20}, and SigFox~\cite{RLORMM18}, which offer a good compromise between coverage, power consumption, and throughput.
However, as \gls{iot} deployments expand into rural regions, \glspl{lpwan} face several limitations.
First, \glspl{lpwan} rely on existing cellular or dedicated network gateways, which are often unavailable or entirely absent in these regions.
Moreover, while \glspl{lpwan} scale well in dense urban deployments, extending coverage in vast and isolated rural areas is often impractical from both a technical and economic point of view.  Additionally, remote areas may not have access to power/energy sources, which further complicates network deployment~\cite{CGLBK21}.

To address these limitations, recent studies have started to explore the integration of \glspl{lpwan} with \glspl{ntn} to support IoT applications~\cite{giordani2021non}.
 In this solution, generally referred to as NTN-IoT~\cite{VAKSACP22}, the IoT gateway is hosted onboard aerial/space nodes operating from the sky, especially \gls{leo} satellites, to provide very large continuous and autonomous geographical coverage, even in the absence of pre-existing terrestrial infrastructures.
However, the existing literature primarily concentrates on urban areas and smart city use cases (e.g.,~\cite{MNHPMR18}), with limited focus on rural or remote areas, which would benefit the most from this paradigm.
For example, NTN-IoT can facilitate smart agriculture, e.g., precision farming via real-time monitoring of soil conditions, even in areas without cellular coverage.
At sea, NTN-IoT allows maritime surveillance and monitoring, e.g., vessel tracking, weather alerts, and monitoring of fishing activities, when terrestrial signals are unavailable. Furthermore, NTN-IoT ensures resilient, always-on communication during or after disasters, when ground infrastructure may be damaged or inaccessible, to provide coordination of rescue teams, especially in remote or hard-to-reach areas.

The feasibility of direct satellite access from \gls{iot} devices has been explored in~\cite{SPQ22} and~\cite{AP22} using NB-IoT and \gls{lora}, respectively.
Additionally, Long Range–Frequency Hopping Spread Spectrum (LR-FHSS), a physical-layer enhancement of \gls{lora}, has been proposed in~\cite{BTAWX21} to improve the uplink performance in \gls{ntn} scenarios.
The \gls{3gpp} is also supporting this framework, and NB-IoT has been considered the leading candidate for \gls{ntn}-\gls{iot} deployments, as explored in~\cite{GVMC20}.
However, preliminary performance evaluations suggested that NB-IoT may not be optimal, especially in terms of link budget and delay~\cite{WTGAZ22}.
Rather, the flexibility of LoRa permits to increase the communication range beyond the limits of NB-IoT, and turns out to be the best approach for LEO satellites.
However, most of these studies focus primarily on the link layer, and introduce many assumptions and simplifications across the network stack.

In this work, we investigate the feasibility and end-to-end performance of \gls{ntn}-IoT based on \gls{lora}. Specifically, we analyze the data rate and \gls{prr} of the access link between IoT end nodes and a \gls{leo} satellite gateway, as a function of the altitude of the satellite, the antenna configuration, and the amount of traffic at the application.
To do so, we use ns-3, one of the most accurate tools for network simulations. 
Although ns-3 includes the \texttt{ns3-LoRa} and \texttt{ns3-NTN} modules to simulate standalone LoRa and NTN scenarios, respectively, it does not natively integrate the two.
To bridge this gap, we developed a new simulation module, called \texttt{ns3-LoRa-NTN}, that integrates the 3GPP \gls{ntn} channel model into the LoRaWAN protocol stack.
Our results show that, despite the more severe propagation environment in satellite networks, the flexibility of LoRa allows ground users to adjust the \gls{sf}, and to communicate with LEO gateways over long distances.

The remainder of the paper is organized as follows. In~\cref{sec:model} we introduce our system model and the \gls{lora} specifications. In~\cref{sub:framework} we present our ns-3 simulation framework. In~\cref{sec:eval} we discuss the simulation results.
Finally, conclusions are drawn in~\cref{sec:conclusions}.

\section{System Model}
\label{sec:model}

In this section we describe our scenario (\cref{sub:scenario}), review the main features of LoRa for IoT (\cref{sub:lora}), and present our channel model (\cref{sec:channel}).


\subsection{Scenario}
\label{sub:scenario}


In this work, we consider a network with $N$ \gls{lora} \glspl{ed}, uniformly distributed across a circular area of size $A_c$.
A single \gls{leo} satellite, acting as a \gls{lora} gateway, is deployed at an altitude $h$, and communicates directly with the \glspl{ed} through a directional antenna with a beamwidth $\theta$. 
The antenna footprint on the ground forms a circular area of radius $R_c$ given by
\begin{equation}
    R_c = \tan \left( {{\theta}/{2}} \right) \cdot h.
    \label{eq:R_c}
\end{equation}
The resulting service area is
\begin{equation}
    A_c = R^2_c \pi.
    \label{eq:A_c}
\end{equation}
For a given \gls{ed} spatial density $\rho_d$, the total number of \glspl{ed} in the service area is computed as
\begin{equation}
    N = \rho_d A_c.
\end{equation}
The communication distance between an \gls{ed} and the LEO satellite is determined by the slant range $d$, which depends on the ED position within the service area. As described in~\cite{38811}, this distance can be derived as
\begin{equation}
d = \sqrt{R^2_E \sin^2(\alpha) + h^2 + 2 h R_E} - R_E \sin(\alpha),
\label{eq:distance}
\end{equation}
where $R_E = 6371~\text{km}$ is the Earth's radius, and $\alpha$ is the elevation angle between the \gls{ed} and the satellite.

\subsection{LoRa}
\label{sub:lora}
In this study, we consider the case in which \glspl{ed} communicate with the LEO satellite gateway using the \gls{lora} technology. 
Specifically, \gls{lora} is a proprietary physical layer modulation technique based on \gls{css}, which offers long-range, low-power connectivity, making it suitable for massive \gls{iot} deployments.

In \gls{css}, data packets are modulated using a chirp signal/waveform whose frequency increases (up-chirp) or decreases (down-chirp) linearly over a fixed bandwidth $B$.
The rate at which the frequency of the signal changes during the symbol duration, i.e., the chirp rate $k$, is given by
\begin{equation}
k = {B}/2^\text{SF},
\end{equation}
where $\text{SF} \in \{7, \ldots, 12\}$ represents the \acrfull{sf}~\cite{Hou22Lora}. Therefore, the SF formally defines the number of chirps encoded per symbol, and directly determines the symbol duration and the data rate~\cite{Sun22Recent}.
As the SF increases, the symbol duration increases and the data rate decreases. At the same time, the receiver sensitivity improves, and so the range of the signal increases~\cite{Voigt17Mitigating}.
This is especially important in the NTN-IoT scenario, where \glspl{ed} tend to increase the SF to maximize the communication range.
Assuming ${B}=125$~kHz, a PHY payload size $p_s$ of $32$~bytes, explicit header mode, and code rate equal to 2, 
the data rates (and \gls{dr} index), sensitivity, and time on air vary significantly based on the SF, as reported in~\cref{tab:lora_sf_parameters}.

A key feature of the \gls{lora} modulation is that SFs are pseudo-orthogonal, so that multiple signals at different data rates (i.e., using different SFs) on the same channel (i.e., using the same time/frequency resources) can be received simultaneously. 
In practice, successful decoding is still possible as long as the desired signal has a sufficiently higher power (generally about 6 dB higher) than the interfering signals~\cite{Croce2018Impact}.

Formally, LoRa operates at the physical layer, and is usually combined with LoRaWAN for the implementation of the rest of the protocol stack, especially the \gls{mac} layer. 
While \gls{lora} transceivers available today can operate in licensed bands between 137 MHz and 1020 MHz, they generally use the sub-GHz \gls{ism} bands (i.e., 433 MHz and 868 MHz in Europe, and 915 MHz in North America)~\cite{Voigt17Mitigating}.

\begin{table}[t]
\centering
\caption{LoRa parameters for different \glspl{sf}, with $B=125$~kHz, $p_s = 32$~bytes, explicit header mode, and code rate equal to 2.}
\fontsize{8}{10}\selectfont
\begin{tabular}{c c c c c}
\toprule
{SF} & {DR} & {Data rate [kbit/s]} & {Sensitivity [dBm]} & {Time on air [ms]} \\
\midrule
7  & 5 & 5.470 & -130.0 & 74 \\
8  & 4 & 3.125 & -132.5 & 136 \\
9  & 3 & 1.760 & -135.0 & 247 \\
10 & 2 & 0.980 & -137.5 & 493 \\
11 & 1 & 0.440 & -140.0 & 888 \\
12 & 0 & 0.250 & -142.5 & 1777 \\
\bottomrule
\end{tabular}
\label{tab:lora_sf_parameters}
\vskip -0.4cm
\end{table}

\subsection{Channel Model}
\label{sec:channel}
The channel between the \glspl{ed} and the LEO satellite is modeled based on the \gls{3gpp} TR 38.811 specifications~\cite{38811}. 
The path loss ($PL$) consists of the \gls{fspl} and additional attenuation components, and is derived~as
\begin{equation}
PL = 20\log_{10}\left(\frac{4\pi d}{\lambda}\right) + L_{a} + CL + SF,
\end{equation}
where $d$ is the distance, $\lambda$ is the wavelength, $L_{a}$ is the atmospheric loss, $SF$ is the shadowing, and $CL$ is the clutter loss that accounts for building obstructions.
In \gls{los} conditions, this latter term is typically negligible, and set to 0~dB~\cite{38811}.
The shadowing is modeled as a zero-mean log-normal random variable with variance $\sigma_s^2$, where $\sigma_s$ depends on the environment, elevation angle, and other link-specific factors. 

The atmospheric loss is due to several contributions, such as gas attenuation, rain and cloud attenuation, tropospheric scintillation,  and ionospheric scintillation~\cite{38811}.
However, as \gls{lora} mainly operates in sub-GHz \gls{ism} bands, only the latter component has a significant impact and is included in our model.
This term is computed according to the Gigahertz Scintillation Model~\cite{ITU2012}, which is valid only for geographical regions with a maximum latitude of 20$^\circ$.
Hence, $L_a$ is given~by
\begin{equation}
    L_a = \left(\frac{f_c}{4}\right)^{-1.5} \frac{P_f \text{ (4 GHz)}}{\sqrt{2}},
\end{equation}
where $f_c$ is the carrier frequency, and $P_f$ (4 GHz) is a scaling factor representing the ionospheric attenuation level at 99\% of the time observed in Hong Kong between March 1977 and March 1978 at a frequency of 4 GHz~\cite{38811}.

The received power $P_{rx}$ (in dB) can be expressed as
\begin{equation}
    P_{rx} = P_{tx} + G_{tx} + G_{rx} - PL,
    \label{eq:P_rx}
\end{equation}
where $P_{tx}$ is the transmit power (in dB), and $G_{tx}$ and $G_{rx}$ are the transmit and receive antenna gains (in dBi), respectively.

At the beginning of the simulation, each \gls{ed} is assigned a SF according to the gateway sensitivity. Specifically, we assign the lowest possible SF
that guarantees that the received power $P_{rx}$ in Eq.~\eqref{eq:P_rx} is higher than the gateway sensitivity.

\section{Simulation Framework}
\label{sub:framework}

Ns-3 is one of the most cited and advanced discrete-event network simulators. Unlike link-level simulators, which often simplify network protocols to reduce the computational complexity, ns-3 incorporates accurate models of the entire stack, enabling full-scale end-to-end simulations. 
In this paper, we develop a new module called \texttt{ns3-LoRa-NTN} to simulate NTN-IoT networks based on LoRa.
This module combines the LoRa PHY/MAC stack from \texttt{ns3-LoRa} with the satellite channel and antenna models from \texttt{ns3-NTN}, adding support for end-to-end NTN-IoT simulations that were not possible with either module alone.
Specifically, the \texttt{ns3-NTN}\footnote{\url{https://gitlab.com/mattiasandri/ns-3-ntn/-/tree/ntn-dev}} module~\cite{sandri2023implementation} is an open-source extension of the ns-3 simulator, developed to model \gls{ntn} communication based on 3GPP Release 17 specifications.
The core of the module is the implementation of the 3GPP TR 38.811 satellite channel model~\cite{38811}, which extends the terrestrial channel model defined in TR~38.901~\cite{38901}.
The model supports multiple propagation environments (e.g., urban, suburban, rural), each associated with specific path loss, \gls{los} probability, and fast fading profiles.
It accounts for atmospheric absorption, modeled using a simplified version of the ITU-R P.676 specification~\cite{itup676}, and scintillation effects, which are derived from tropospheric and ionospheric models~\cite{ITU2012}.
In line with the 3GPP specifications, \texttt{ns3-NTN} includes antenna models for circular aperture antennas for satellite terminals, and \gls{upa} and \gls{vsat} antennas for \glspl{ed}~\cite{38811}.
Moreover, the module implements a Geocentric Cartesian coordinate (ECEF) system, 
to consider the Earth's curvature, as well as the elevation angle, to represent the position of a node. The x-y plane defines the equatorial plane, with the x-axis pointing at $0^\circ$ longitude, the y-axis pointing at $90^\circ$ longitude, and the z-axis pointing at the geographical North Pole.
Finally, it extends the benchmark ns-3 code to model the propagation delay between end nodes (which is typically negligible for terrestrial links), and a Timing Advance (TA) mechanism to compensate for this delay during scheduling~\cite{38214}.
\gls{rrc} and \gls{harq} timers have also been properly adjusted to account for the long propagation delay~\cite{38821}.
  
The \texttt{ns3-LoRa}\footnote{\url{https://github.com/signetlabdei/lorawan}} module is an open-source ns-3 framework to simulate \gls{lora} networks at various levels~\cite{magrin2017performance}.
It supports the simulation of class A \glspl{ed}, in which each \gls{ed} transmits
application packets on the wireless channel asynchronously.
After each uplink transmission, the node opens up at most two reception windows, waiting for any command or data packet returned by the \gls{ns}.
The system takes into account channel interference, and considers a packet delivered successfully only if the equalized
interfering power is below the channel rejection parameter defined in~\cite{Goursaud15}.
Additionally, it integrates both energy harvesting and consumption models, and allows for flexible network configuration, such as enabling or disabling the duty cycle restriction at the gateway.
Furthermore, this module enables \glspl{ed} to freely select the \gls{sf} and channel for the reception windows, with no restrictions on the values specified by the LoRaWAN standard.
However,  \texttt{ns3-LoRa} does not support simulations in the \gls{ntn} scenario, which motivates our research work in this paper. 

The \texttt{ns3-LoRa-NTN} module is the first open-source module able to simulate an NTN-IoT network, where terrestrial \glspl{ed} communicate with a satellite gateway, thereby expanding the scope of traditional \gls{lpwan} applications.

\section{Performance Evaluation}
\label{sec:eval}
In this section, we first describe our simulation setup and parameters. Then, we present some results to validate the NTN-IoT paradigm using LoRa.

\begin{table}[t]
  \caption{Simulation parameters.}
  \label{tab:parameters}
  \footnotesize
  \centering
  \renewcommand{\arraystretch}{1.2}
  \begin{tabular}{|l|l|}
    \hline
    {Parameter} & {Value} \\ 
    \hline
    Satellite altitude ($h$) [km] & [200, \dots, 700]\\
    Transmit power ($P_{tx})$ [dBm] & 14 \\
    Bandwidth ($B$) [kHz] & 125 \\
    Carrier frequency ($f_c$) [MHz] & 868 \\
    \gls{ed} density ($\rho_d$) [\glspl{ed}/km$^2$] & 0.01 \\
    Antenna beamwidth ($\theta$) [$^\circ$] & \{5, 10, 15\} \\
    Total antenna gain ($G_{tx} + G_{rx}$) [dBi] & \{5, 10\} \\
    Transmission periodicity ($p$) [s] & \{60, 30, 10\} \\
    PHY payload size ($p_s$) [bytes] & 32 \\
    Earth's radius ($R_E$) [km] & 6371 \\
    \hline
  \end{tabular}
  \vskip -0.5cm
\end{table}

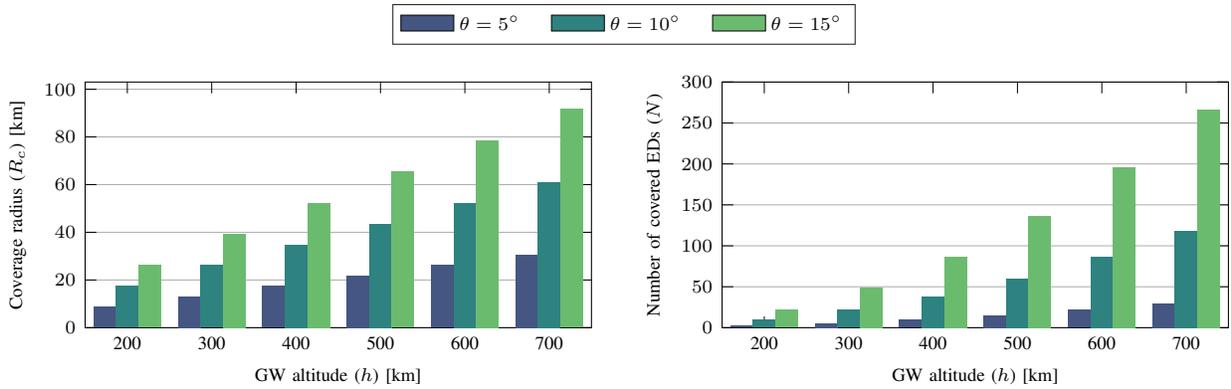
\begin{figure*}[t!]
\centering
%
%

\definecolor{darkslateblue6886129}{RGB}{68,86,129}
\definecolor{mediumseagreen107187110}{RGB}{107,187,110}
\definecolor{seagreen46130127}{RGB}{46,130,127}

\begin{tikzpicture}
\pgfplotsset{every tick label/.append style={font=\scriptsize}}

\pgfplotsset{compat=1.11,
	/pgfplots/ybar legend/.style={
		/pgfplots/legend image code/.code={%
			\draw[##1,/tikz/.cd,yshift=-0.25em]
			(0cm,0cm) rectangle (20pt,0.6em);},
	},
}

\begin{axis}[%
width=0,
height=0,
at={(0,0)},
scale only axis,
xmin=0,
xmax=0,
xtick={},
ymin=0,
ymax=0,
ytick={},
axis background/.style={fill=white},
legend style={legend cell align=left,
              align=center,
              draw=white!15!black,
              at={(0.5, 1.3)},
              anchor=center,
              /tikz/every even column/.append style={column sep=1em}},
legend columns=3,
]
\addplot[ybar,ybar legend,draw=black,fill=darkslateblue6886129,line width=0.08pt]
table[row sep=crcr]{%
	0	0\\
};
\addlegendentry{$\theta=5^\circ$}

\addplot[ybar,ybar legend,draw=black,fill=seagreen46130127,line width=0.08pt]
table[row sep=crcr]{%
	0	0\\
};
\addlegendentry{$\theta=10^\circ$}

\addplot[ybar,ybar legend,draw=black,fill=mediumseagreen107187110,line width=0.08pt]
table[row sep=crcr]{%
	0	0\\
};
\addlegendentry{$\theta=15^\circ$}

\end{axis}
\end{tikzpicture}%
\vskip 0.3cm
    {
        \label{fig:coverage_radius}
\begin{tikzpicture}

\definecolor{darkgray176}{RGB}{176,176,176}
\definecolor{darkslateblue6886129}{RGB}{68,86,129}
\definecolor{darkslategray66}{RGB}{66,66,66}
\definecolor{lightgray204}{RGB}{204,204,204}
\definecolor{mediumseagreen107187110}{RGB}{107,187,110}
\definecolor{seagreen46130127}{RGB}{46,130,127}

\pgfplotsset{compat=1.11,
	/pgfplots/ybar legend/.style={
		/pgfplots/legend image code/.code={%
			\draw[##1,/tikz/.cd,yshift=-0.25em]
			(0cm,0cm) rectangle (20pt,0.6em);},
	},
}

\begin{axis}[
width = \textwidth/2.2,
height = 4.85cm,
legend cell align={left},
legend columns=3,
legend style={
  fill opacity=0.8,
  draw opacity=1,
  text opacity=1,
  at={(0.5,1.18)},
  anchor=north,
  draw=lightgray204,
  /tikz/every even column/.append style={column sep=0.5em}
},
tick pos=both,
unbounded coords=jump,
x grid style={darkgray176},
xlabel={GW altitude ($h$) [km]},
xmin=-0.5, xmax=5.5,
xtick style={color=black},
xtick={0,1,2,3,4,5,6},
xticklabels={200, 300, 400, 500, 600, 700},
y grid style={darkgray176},
ylabel={Coverage radius ($R_c$) [km]},
ymajorgrids,
ymin=0, ymax=103,
ytick style={color=black},
ytick={0,20,40,60,80,100,120},
yticklabels={
  \(\displaystyle {0}\),
  \(\displaystyle {20}\),
  \(\displaystyle {40}\),
  \(\displaystyle {60}\),
  \(\displaystyle {80}\),
  \(\displaystyle {100}\),
  \(\displaystyle {120}\)
}
]
\draw[draw=none,fill=darkslateblue6886129] (axis cs:-0.4,0) rectangle (axis cs:-0.133333333333333,8.72664625997165);

\draw[draw=none,fill=darkslateblue6886129] (axis cs:0.6,0) rectangle (axis cs:0.866666666666667,13.0899693899575);
\draw[draw=none,fill=darkslateblue6886129] (axis cs:1.6,0) rectangle (axis cs:1.86666666666667,17.4532925199433);
\draw[draw=none,fill=darkslateblue6886129] (axis cs:2.6,0) rectangle (axis cs:2.86666666666667,21.8166156499291);
\draw[draw=none,fill=darkslateblue6886129] (axis cs:3.6,0) rectangle (axis cs:3.86666666666667,26.1799387799149);
\draw[draw=none,fill=darkslateblue6886129] (axis cs:4.6,0) rectangle (axis cs:4.86666666666667,30.5432619099008);
\draw[draw=none,fill=darkslateblue6886129] (axis cs:5.6,0) rectangle (axis cs:5.86666666666667,34.9065850398866);
\draw[draw=none,fill=seagreen46130127] (axis cs:-0.133333333333333,0) rectangle (axis cs:0.133333333333333,17.4532925199433);

\draw[draw=none,fill=seagreen46130127] (axis cs:0.866666666666667,0) rectangle (axis cs:1.13333333333333,26.1799387799149);
\draw[draw=none,fill=seagreen46130127] (axis cs:1.86666666666667,0) rectangle (axis cs:2.13333333333333,34.9065850398866);
\draw[draw=none,fill=seagreen46130127] (axis cs:2.86666666666667,0) rectangle (axis cs:3.13333333333333,43.6332312998582);
\draw[draw=none,fill=seagreen46130127] (axis cs:3.86666666666667,0) rectangle (axis cs:4.13333333333333,52.3598775598299);
\draw[draw=none,fill=seagreen46130127] (axis cs:4.86666666666667,0) rectangle (axis cs:5.13333333333333,61.0865238198015);
\draw[draw=none,fill=seagreen46130127] (axis cs:5.86666666666667,0) rectangle (axis cs:6.13333333333333,69.8131700797732);
\draw[draw=none,fill=mediumseagreen107187110] (axis cs:0.133333333333333,0) rectangle (axis cs:0.4,26.1799387799149);

\draw[draw=none,fill=mediumseagreen107187110] (axis cs:1.13333333333333,0) rectangle (axis cs:1.4,39.2699081698724);
\draw[draw=none,fill=mediumseagreen107187110] (axis cs:2.13333333333333,0) rectangle (axis cs:2.4,52.3598775598299);
\draw[draw=none,fill=mediumseagreen107187110] (axis cs:3.13333333333333,0) rectangle (axis cs:3.4,65.4498469497874);
\draw[draw=none,fill=mediumseagreen107187110] (axis cs:4.13333333333333,0) rectangle (axis cs:4.4,78.5398163397448);
\draw[draw=none,fill=mediumseagreen107187110] (axis cs:5.13333333333333,0) rectangle (axis cs:5.4,91.6297857297023);
\draw[draw=none,fill=mediumseagreen107187110] (axis cs:6.13333333333333,0) rectangle (axis cs:6.4,104.71975511966);
\addplot [line width=0.72pt, darkslategray66, forget plot]
table {%
-0.266666666666667 nan
-0.266666666666667 nan
};
\addplot [line width=0.72pt, darkslategray66, forget plot]
table {%
0.733333333333333 nan
0.733333333333333 nan
};
\addplot [line width=0.72pt, darkslategray66, forget plot]
table {%
1.73333333333333 nan
1.73333333333333 nan
};
\addplot [line width=0.72pt, darkslategray66, forget plot]
table {%
2.73333333333333 nan
2.73333333333333 nan
};
\addplot [line width=0.72pt, darkslategray66, forget plot]
table {%
3.73333333333333 nan
3.73333333333333 nan
};
\addplot [line width=0.72pt, darkslategray66, forget plot]
table {%
4.73333333333333 nan
4.73333333333333 nan
};
\addplot [line width=0.72pt, darkslategray66, forget plot]
table {%
5.73333333333333 nan
5.73333333333333 nan
};
\addplot [line width=0.72pt, darkslategray66, forget plot]
table {%
0 nan
0 nan
};
\addplot [line width=0.72pt, darkslategray66, forget plot]
table {%
1 nan
1 nan
};
\addplot [line width=0.72pt, darkslategray66, forget plot]
table {%
2 nan
2 nan
};
\addplot [line width=0.72pt, darkslategray66, forget plot]
table {%
3 nan
3 nan
};
\addplot [line width=0.72pt, darkslategray66, forget plot]
table {%
4 nan
4 nan
};
\addplot [line width=0.72pt, darkslategray66, forget plot]
table {%
5 nan
5 nan
};
\addplot [line width=0.72pt, darkslategray66, forget plot]
table {%
6 nan
6 nan
};
\addplot [line width=0.72pt, darkslategray66, forget plot]
table {%
0.266666666666667 nan
0.266666666666667 nan
};
\addplot [line width=0.72pt, darkslategray66, forget plot]
table {%
1.26666666666667 nan
1.26666666666667 nan
};
\addplot [line width=0.72pt, darkslategray66, forget plot]
table {%
2.26666666666667 nan
2.26666666666667 nan
};
\addplot [line width=0.72pt, darkslategray66, forget plot]
table {%
3.26666666666667 nan
3.26666666666667 nan
};
\addplot [line width=0.72pt, darkslategray66, forget plot]
table {%
4.26666666666667 nan
4.26666666666667 nan
};
\addplot [line width=0.72pt, darkslategray66, forget plot]
table {%
5.26666666666667 nan
5.26666666666667 nan
};
\addplot [line width=0.72pt, darkslategray66, forget plot]
table {%
6.26666666666667 nan
6.26666666666667 nan
};
\end{axis}

\end{tikzpicture}
    }
    {
        \label{fig:ed_covered}
\begin{tikzpicture}

\definecolor{darkgray176}{RGB}{176,176,176}
\definecolor{darkslateblue6886129}{RGB}{68,86,129}
\definecolor{darkslategray66}{RGB}{66,66,66}
\definecolor{lightgray204}{RGB}{204,204,204}
\definecolor{mediumseagreen107187110}{RGB}{107,187,110}
\definecolor{seagreen46130127}{RGB}{46,130,127}

\pgfplotsset{compat=1.11,
	/pgfplots/ybar legend/.style={
		/pgfplots/legend image code/.code={%
			\draw[##1,/tikz/.cd,yshift=-0.25em]
			(0cm,0cm) rectangle (20pt,0.6em);},
	},
}

\begin{axis}[
width = \textwidth/2.2,
height = 4.85cm,
legend cell align={left},
legend columns=3,
legend style={
  fill opacity=0.8,
  draw opacity=1,
  text opacity=1,
  at={(0.5,1.18)},
  anchor=north,
  draw=lightgray204,
  /tikz/every even column/.append style={column sep=0.5em}
},
tick pos=both,
unbounded coords=jump,
x grid style={darkgray176},
xlabel={GW altitude ($h$) [km]},
xmin=-0.5, xmax=5.5,
xtick style={color=black},
xtick={0,1,2,3,4,5,6},
xticklabels={200, 300, 400, 500, 600, 700},
y grid style={darkgray176},
ylabel={Number of covered EDs ($N$)},
ymajorgrids,
ymin=0, ymax=300,
ytick style={color=black},
ytick={0,50,100,150,200,250,300}
]
\draw[draw=none,fill=darkslateblue6886129] (axis cs:-0.4,0) rectangle (axis cs:-0.133333333333333,2.39549950334835);

\draw[draw=none,fill=darkslateblue6886129] (axis cs:0.6,0) rectangle (axis cs:0.866666666666667,5.38987388253378);
\draw[draw=none,fill=darkslateblue6886129] (axis cs:1.6,0) rectangle (axis cs:1.86666666666667,9.58199801339339);
\draw[draw=none,fill=darkslateblue6886129] (axis cs:2.6,0) rectangle (axis cs:2.86666666666667,14.9718718959272);
\draw[draw=none,fill=darkslateblue6886129] (axis cs:3.6,0) rectangle (axis cs:3.86666666666667,21.5594955301351);
\draw[draw=none,fill=darkslateblue6886129] (axis cs:4.6,0) rectangle (axis cs:4.86666666666667,29.3448689160173);
\draw[draw=none,fill=seagreen46130127] (axis cs:-0.133333333333333,0) rectangle (axis cs:0.133333333333333,9.61863464225903);

\draw[draw=none,fill=seagreen46130127] (axis cs:0.866666666666667,0) rectangle (axis cs:1.13333333333333,21.6419279450828);
\draw[draw=none,fill=seagreen46130127] (axis cs:1.86666666666667,0) rectangle (axis cs:2.13333333333333,38.4745385690361);
\draw[draw=none,fill=seagreen46130127] (axis cs:2.86666666666667,0) rectangle (axis cs:3.13333333333333,60.1164665141189);
\draw[draw=none,fill=seagreen46130127] (axis cs:3.86666666666667,0) rectangle (axis cs:4.13333333333333,86.5677117803313);
\draw[draw=none,fill=seagreen46130127] (axis cs:4.86666666666667,0) rectangle (axis cs:5.13333333333333,117.828274367673);
\draw[draw=none,fill=mediumseagreen107187110] (axis cs:0.133333333333333,0) rectangle (axis cs:0.4,21.7805112229428);

\draw[draw=none,fill=mediumseagreen107187110] (axis cs:1.13333333333333,0) rectangle (axis cs:1.4,49.0061502516213);
\draw[draw=none,fill=mediumseagreen107187110] (axis cs:2.13333333333333,0) rectangle (axis cs:2.4,87.1220448917713);
\draw[draw=none,fill=mediumseagreen107187110] (axis cs:3.13333333333333,0) rectangle (axis cs:3.4,136.128195143393);
\draw[draw=none,fill=mediumseagreen107187110] (axis cs:4.13333333333333,0) rectangle (axis cs:4.4,196.024601006485);
\draw[draw=none,fill=mediumseagreen107187110] (axis cs:5.13333333333333,0) rectangle (axis cs:5.4,266.81126248105);
\addplot [line width=0.72pt, darkslategray66, forget plot]
table {%
-0.266666666666667 nan
-0.266666666666667 nan
};
\addplot [line width=0.72pt, darkslategray66, forget plot]
table {%
0.733333333333333 nan
0.733333333333333 nan
};
\addplot [line width=0.72pt, darkslategray66, forget plot]
table {%
1.73333333333333 nan
1.73333333333333 nan
};
\addplot [line width=0.72pt, darkslategray66, forget plot]
table {%
2.73333333333333 nan
2.73333333333333 nan
};
\addplot [line width=0.72pt, darkslategray66, forget plot]
table {%
3.73333333333333 nan
3.73333333333333 nan
};
\addplot [line width=0.72pt, darkslategray66, forget plot]
table {%
4.73333333333333 nan
4.73333333333333 nan
};
\addplot [line width=0.72pt, darkslategray66, forget plot]
table {%
0 nan
0 nan
};
\addplot [line width=0.72pt, darkslategray66, forget plot]
table {%
1 nan
1 nan
};
\addplot [line width=0.72pt, darkslategray66, forget plot]
table {%
2 nan
2 nan
};
\addplot [line width=0.72pt, darkslategray66, forget plot]
table {%
3 nan
3 nan
};
\addplot [line width=0.72pt, darkslategray66, forget plot]
table {%
4 nan
4 nan
};
\addplot [line width=0.72pt, darkslategray66, forget plot]
table {%
5 nan
5 nan
};
\addplot [line width=0.72pt, darkslategray66, forget plot]
table {%
0.266666666666667 nan
0.266666666666667 nan
};
\addplot [line width=0.72pt, darkslategray66, forget plot]
table {%
1.26666666666667 nan
1.26666666666667 nan
};
\addplot [line width=0.72pt, darkslategray66, forget plot]
table {%
2.26666666666667 nan
2.26666666666667 nan
};
\addplot [line width=0.72pt, darkslategray66, forget plot]
table {%
3.26666666666667 nan
3.26666666666667 nan
};
\addplot [line width=0.72pt, darkslategray66, forget plot]
table {%
4.26666666666667 nan
4.26666666666667 nan
};
\addplot [line width=0.72pt, darkslategray66, forget plot]
table {%
5.26666666666667 nan
5.26666666666667 nan
};
\end{axis}
\end{tikzpicture}
    }
\caption{Coverage radius (left) and the number of covered \glspl{ed} (right) vs. $\theta$ and~$h$, with $p=60$ s.}
\label{fig:coverage}
\vskip -0.3cm
\end{figure*}

\begin{figure*}[t!]
\centering
\subfloat[DR distribution, for $\theta=5^\circ$.]
{
    \label{fig:dr_distribution_1}
\begin{tikzpicture}

\definecolor{darkgray176}{RGB}{176,176,176}
\definecolor{darkslategray38}{RGB}{38,38,38}

\begin{axis}[
width = \textwidth/3.5,
height = 4.85cm,
colorbar,
colorbar style={ytick={0,0.2,0.4,0.6,0.8,1},yticklabels={
  \(\displaystyle {0.0}\),
  \(\displaystyle {0.2}\),
  \(\displaystyle {0.4}\),
  \(\displaystyle {0.6}\),
  \(\displaystyle {0.8}\),
  \(\displaystyle {1.0}\)
},ylabel={Data rate allocation [\%]}},
colormap={mymap}{[1pt]
  rgb(0pt)=(0,0,1);
  rgb(1pt)=(0,1,0.5)
},
point meta max=1,
point meta min=0,
tick pos=both,
x grid style={darkgray176},
xlabel={GW altitude ($h$) [km]},
xmin=0, xmax=6,
xtick style={color=black},
xtick={0.5,1.5,2.5,3.5,4.5,5.5},
xticklabels={200,300,400,500,600,700},
y dir=reverse,
y grid style={darkgray176},
ylabel={\acrfull{dr} index},
ymin=0, ymax=6,
ytick style={color=black},
ytick={0.5,1.5,2.5,3.5,4.5,5.5},
yticklabels={0,1,2,3,4,5}
]
\addplot graphics [includegraphics cmd=\pgfimage,xmin=0, xmax=6, ymin=6, ymax=0] {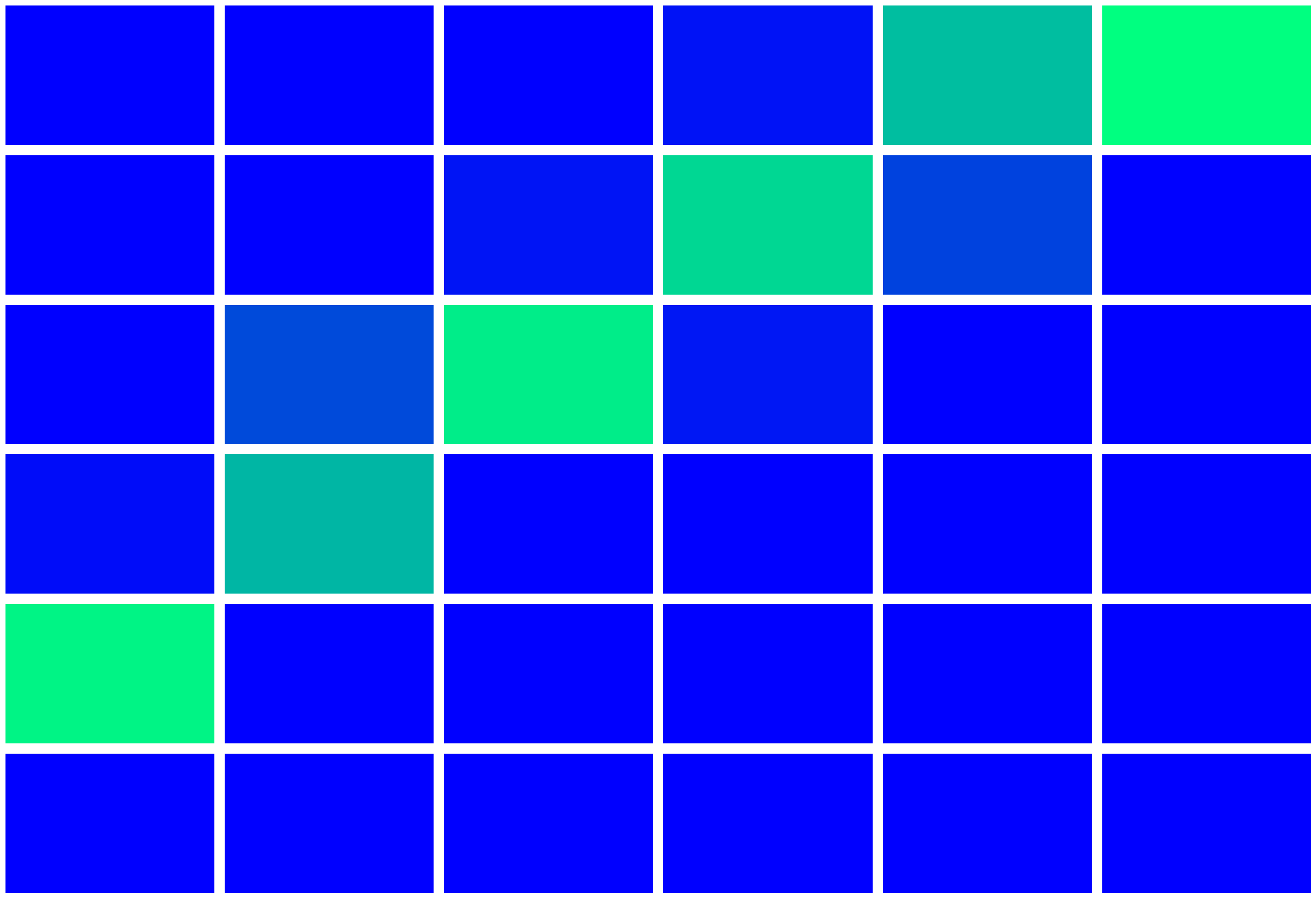};
\draw (axis cs:0.5,0.5) node[
  scale=0.6,
  text=white,
  rotate=0.0
]{0.00};
\draw (axis cs:1.5,0.5) node[
  scale=0.6,
  text=white,
  rotate=0.0
]{0.00};
\draw (axis cs:2.5,0.5) node[
  scale=0.6,
  text=white,
  rotate=0.0
]{0.00};
\draw (axis cs:3.5,0.5) node[
  scale=0.6,
  text=white,
  rotate=0.0
]{0.07};
\draw (axis cs:4.5,0.5) node[
  scale=0.6,
  text=white,
  rotate=0.0
]{0.74};
\draw (axis cs:5.5,0.5) node[
  scale=0.6,
  text=darkslategray38,
  rotate=0.0
]{0.99};
\draw (axis cs:0.5,1.5) node[
  scale=0.6,
  text=white,
  rotate=0.0
]{0.00};
\draw (axis cs:1.5,1.5) node[
  scale=0.6,
  text=white,
  rotate=0.0
]{0.00};
\draw (axis cs:2.5,1.5) node[
  scale=0.6,
  text=white,
  rotate=0.0
]{0.08};
\draw (axis cs:3.5,1.5) node[
  scale=0.6,
  text=darkslategray38,
  rotate=0.0
]{0.84};
\draw (axis cs:4.5,1.5) node[
  scale=0.6,
  text=white,
  rotate=0.0
]{0.26};
\draw (axis cs:5.5,1.5) node[
  scale=0.6,
  text=white,
  rotate=0.0
]{0.01};
\draw (axis cs:0.5,2.5) node[
  scale=0.6,
  text=white,
  rotate=0.0
]{0.00};
\draw (axis cs:1.5,2.5) node[
  scale=0.6,
  text=white,
  rotate=0.0
]{0.29};
\draw (axis cs:2.5,2.5) node[
  scale=0.6,
  text=darkslategray38,
  rotate=0.0
]{0.92};
\draw (axis cs:3.5,2.5) node[
  scale=0.6,
  text=white,
  rotate=0.0
]{0.09};
\draw (axis cs:4.5,2.5) node[
  scale=0.6,
  text=white,
  rotate=0.0
]{0.00};
\draw (axis cs:5.5,2.5) node[
  scale=0.6,
  text=white,
  rotate=0.0
]{0.00};
\draw (axis cs:0.5,3.5) node[
  scale=0.6,
  text=white,
  rotate=0.0
]{0.05};
\draw (axis cs:1.5,3.5) node[
  scale=0.6,
  text=white,
  rotate=0.0
]{0.71};
\draw (axis cs:2.5,3.5) node[
  scale=0.6,
  text=white,
  rotate=0.0
]{0.00};
\draw (axis cs:3.5,3.5) node[
  scale=0.6,
  text=white,
  rotate=0.0
]{0.00};
\draw (axis cs:4.5,3.5) node[
  scale=0.6,
  text=white,
  rotate=0.0
]{0.00};
\draw (axis cs:5.5,3.5) node[
  scale=0.6,
  text=white,
  rotate=0.0
]{0.00};
\draw (axis cs:0.5,4.5) node[
  scale=0.6,
  text=darkslategray38,
  rotate=0.0
]{0.95};
\draw (axis cs:1.5,4.5) node[
  scale=0.6,
  text=white,
  rotate=0.0
]{0.00};
\draw (axis cs:2.5,4.5) node[
  scale=0.6,
  text=white,
  rotate=0.0
]{0.00};
\draw (axis cs:3.5,4.5) node[
  scale=0.6,
  text=white,
  rotate=0.0
]{0.00};
\draw (axis cs:4.5,4.5) node[
  scale=0.6,
  text=white,
  rotate=0.0
]{0.00};
\draw (axis cs:5.5,4.5) node[
  scale=0.6,
  text=white,
  rotate=0.0
]{0.00};
\draw (axis cs:0.5,5.5) node[
  scale=0.6,
  text=white,
  rotate=0.0
]{0.00};
\draw (axis cs:1.5,5.5) node[
  scale=0.6,
  text=white,
  rotate=0.0
]{0.00};
\draw (axis cs:2.5,5.5) node[
  scale=0.6,
  text=white,
  rotate=0.0
]{0.00};
\draw (axis cs:3.5,5.5) node[
  scale=0.6,
  text=white,
  rotate=0.0
]{0.00};
\draw (axis cs:4.5,5.5) node[
  scale=0.6,
  text=white,
  rotate=0.0
]{0.00};
\draw (axis cs:5.5,5.5) node[
  scale=0.6,
  text=white,
  rotate=0.0
]{0.00};
\end{axis}

\end{tikzpicture}
}
\begin{minipage}[b]{0.64\textwidth}
    \centering
%
%

\definecolor{darkslateblue6886129}{RGB}{68,86,129}
\definecolor{mediumseagreen107187110}{RGB}{107,187,110}
\definecolor{seagreen46130127}{RGB}{46,130,127}

\begin{tikzpicture}
\pgfplotsset{every tick label/.append style={font=\scriptsize}}

\pgfplotsset{compat=1.11,
	/pgfplots/ybar legend/.style={
		/pgfplots/legend image code/.code={%
			\draw[##1,/tikz/.cd,yshift=-0.25em]
			(0cm,0cm) rectangle (20pt,0.6em);},
	},
}

\begin{axis}[%
width=0,
height=0,
at={(0,0)},
scale only axis,
xmin=0,
xmax=0,
xtick={},
ymin=0,
ymax=0,
ytick={},
axis background/.style={fill=white},
legend style={legend cell align=left,
              align=center,
              draw=white!15!black,
              at={(0.5, 1.3)},
              anchor=center,
              /tikz/every even column/.append style={column sep=1em}},
legend columns=3,
]
\addplot[ybar,ybar legend,draw=black,fill=darkslateblue6886129,line width=0.08pt]
table[row sep=crcr]{%
	0	0\\
};
\addlegendentry{$\theta=5^\circ$}

\addplot[ybar,ybar legend,draw=black,fill=seagreen46130127,line width=0.08pt]
table[row sep=crcr]{%
	0	0\\
};
\addlegendentry{$\theta=10^\circ$}

\addplot[ybar,ybar legend,draw=black,fill=mediumseagreen107187110,line width=0.08pt]
table[row sep=crcr]{%
	0	0\\
};
\addlegendentry{$\theta=15^\circ$}

\end{axis}
\end{tikzpicture}%
    \vskip 0.2 cm
    \subfloat[Average data rate.]
    {
        \label{fig:dr_avg_1}
\begin{tikzpicture}

\definecolor{darkgray176}{RGB}{176,176,176}
\definecolor{darkslateblue6886129}{RGB}{68,86,129}
\definecolor{darkslategray66}{RGB}{66,66,66}
\definecolor{lightgray204}{RGB}{204,204,204}
\definecolor{mediumseagreen107187110}{RGB}{107,187,110}
\definecolor{seagreen46130127}{RGB}{46,130,127}

\pgfplotsset{compat=1.11,
	/pgfplots/ybar legend/.style={
		/pgfplots/legend image code/.code={%
			\draw[##1,/tikz/.cd,yshift=-0.25em]
			(0cm,0cm) rectangle (20pt,0.6em);},
	},
}

\begin{axis}[
width = \textwidth/2,
height = 4.85cm,
legend cell align={left},
legend columns=3,
legend style={
  fill opacity=0.8,
  draw opacity=1,
  text opacity=1,
  at={(0.5,1.18)},
  anchor=north,
  draw=lightgray204,
  /tikz/every even column/.append style={column sep=0.5em}
},
tick pos=both,
unbounded coords=jump,
x grid style={darkgray176},
xlabel={GW altitude ($h$) [km]},
xmin=-0.5, xmax=5.5,
xtick style={color=black},
xtick={0,1,2,3,4,5},
xticklabels={200,300,400,500,600,700},
y grid style={darkgray176},
ymajorgrids,
ylabel={Average data rate [kbps]},
ymin=0, ymax=3.2095875,
ytick style={color=black},
ytick={0,0.5,1,1.5,2,2.5,3,3.5},
yticklabels={
  \(\displaystyle {0.0}\),
  \(\displaystyle {0.5}\),
  \(\displaystyle {1.0}\),
  \(\displaystyle {1.5}\),
  \(\displaystyle {2.0}\),
  \(\displaystyle {2.5}\),
  \(\displaystyle {3.0}\),
  \(\displaystyle {3.5}\)
}
]

\draw[draw=none,fill=darkslateblue6886129] (axis cs:-0.4,0) rectangle (axis cs:-0.133333333333333,3.05675);

\draw[draw=none,fill=darkslateblue6886129] (axis cs:0.6,0) rectangle (axis cs:0.866666666666667,1.5338);
\draw[draw=none,fill=darkslateblue6886129] (axis cs:1.6,0) rectangle (axis cs:1.86666666666667,0.938);
\draw[draw=none,fill=darkslateblue6886129] (axis cs:2.6,0) rectangle (axis cs:2.86666666666667,0.476571428571429);
\draw[draw=none,fill=darkslateblue6886129] (axis cs:3.6,0) rectangle (axis cs:3.86666666666667,0.299309523809524);
\draw[draw=none,fill=darkslateblue6886129] (axis cs:4.6,0) rectangle (axis cs:4.86666666666667,0.25098275862069);
\draw[draw=none,fill=seagreen46130127] (axis cs:-0.133333333333333,0) rectangle (axis cs:0.133333333333333,3.01883333333333);

\draw[draw=none,fill=seagreen46130127] (axis cs:0.866666666666667,0) rectangle (axis cs:1.13333333333333,1.5632619047619);
\draw[draw=none,fill=seagreen46130127] (axis cs:1.86666666666667,0) rectangle (axis cs:2.13333333333333,0.934302631578947);
\draw[draw=none,fill=seagreen46130127] (axis cs:2.86666666666667,0) rectangle (axis cs:3.13333333333333,0.484883333333333);
\draw[draw=none,fill=seagreen46130127] (axis cs:3.86666666666667,0) rectangle (axis cs:4.13333333333333,0.298127906976744);
\draw[draw=none,fill=seagreen46130127] (axis cs:4.86666666666667,0) rectangle (axis cs:5.13333333333333,0.251136752136752);
\draw[draw=none,fill=mediumseagreen107187110] (axis cs:0.133333333333333,0) rectangle (axis cs:0.4,2.98585714285714);

\draw[draw=none,fill=mediumseagreen107187110] (axis cs:1.13333333333333,0) rectangle (axis cs:1.4,1.54566836734694);
\draw[draw=none,fill=mediumseagreen107187110] (axis cs:2.13333333333333,0) rectangle (axis cs:2.4,0.927212643678161);
\draw[draw=none,fill=mediumseagreen107187110] (axis cs:3.13333333333333,0) rectangle (axis cs:3.4,0.477654411764706);
\draw[draw=none,fill=mediumseagreen107187110] (axis cs:4.13333333333333,0) rectangle (axis cs:4.4,0.298283163265306);
\draw[draw=none,fill=mediumseagreen107187110] (axis cs:5.13333333333333,0) rectangle (axis cs:5.4,0.252402255639098);
\addplot [line width=0.72pt, darkslategray66, forget plot]
table {%
-0.266666666666667 nan
-0.266666666666667 nan
};
\addplot [line width=0.72pt, darkslategray66, forget plot]
table {%
0.733333333333333 nan
0.733333333333333 nan
};
\addplot [line width=0.72pt, darkslategray66, forget plot]
table {%
1.73333333333333 nan
1.73333333333333 nan
};
\addplot [line width=0.72pt, darkslategray66, forget plot]
table {%
2.73333333333333 nan
2.73333333333333 nan
};
\addplot [line width=0.72pt, darkslategray66, forget plot]
table {%
3.73333333333333 nan
3.73333333333333 nan
};
\addplot [line width=0.72pt, darkslategray66, forget plot]
table {%
4.73333333333333 nan
4.73333333333333 nan
};
\addplot [line width=0.72pt, darkslategray66, forget plot]
table {%
0 nan
0 nan
};
\addplot [line width=0.72pt, darkslategray66, forget plot]
table {%
1 nan
1 nan
};
\addplot [line width=0.72pt, darkslategray66, forget plot]
table {%
2 nan
2 nan
};
\addplot [line width=0.72pt, darkslategray66, forget plot]
table {%
3 nan
3 nan
};
\addplot [line width=0.72pt, darkslategray66, forget plot]
table {%
4 nan
4 nan
};
\addplot [line width=0.72pt, darkslategray66, forget plot]
table {%
5 nan
5 nan
};
\addplot [line width=0.72pt, darkslategray66, forget plot]
table {%
0.266666666666667 nan
0.266666666666667 nan
};
\addplot [line width=0.72pt, darkslategray66, forget plot]
table {%
1.26666666666667 nan
1.26666666666667 nan
};
\addplot [line width=0.72pt, darkslategray66, forget plot]
table {%
2.26666666666667 nan
2.26666666666667 nan
};
\addplot [line width=0.72pt, darkslategray66, forget plot]
table {%
3.26666666666667 nan
3.26666666666667 nan
};
\addplot [line width=0.72pt, darkslategray66, forget plot]
table {%
4.26666666666667 nan
4.26666666666667 nan
};
\addplot [line width=0.72pt, darkslategray66, forget plot]
table {%
5.26666666666667 nan
5.26666666666667 nan
};
\end{axis}

\end{tikzpicture}
    }
    \subfloat[Packet Reception Ratio.]
    {
        \label{fig:prr_1}
\begin{tikzpicture}

\definecolor{darkgray176}{RGB}{176,176,176}
\definecolor{darkslateblue6886129}{RGB}{68,86,129}
\definecolor{darkslategray66}{RGB}{66,66,66}
\definecolor{lightgray204}{RGB}{204,204,204}
\definecolor{mediumseagreen107187110}{RGB}{107,187,110}
\definecolor{seagreen46130127}{RGB}{46,130,127}

\pgfplotsset{compat=1.11,
	/pgfplots/ybar legend/.style={
		/pgfplots/legend image code/.code={%
			\draw[##1,/tikz/.cd,yshift=-0.25em]
			(0cm,0cm) rectangle (20pt,0.6em);},
	},
}

\begin{axis}[
width = \textwidth/2,
height = 4.85cm,
legend cell align={left},
legend columns=3,
legend style={
  fill opacity=0.8,
  draw opacity=1,
  text opacity=1,
  at={(0.5,1.18)},
  anchor=north,
  draw=lightgray204,
  /tikz/every even column/.append style={column sep=0.5em}
},
tick pos=both,
unbounded coords=jump,
x grid style={darkgray176},
xlabel={GW altitude ($h$) [km]},
xmin=-0.5, xmax=5.5,
xtick style={color=black},
xtick={0,1,2,3,4,5},
xticklabels={200,300,400,500,600,700},
y grid style={darkgray176},
ylabel={Packet Reception Ratio (PRR)},
ymajorgrids,
ymin=0, ymax=1.05,
ytick style={color=black},
ytick={0,0.2,0.4,0.6,0.8,1,1.2},
yticklabels={
  \(\displaystyle {0.0}\),
  \(\displaystyle {0.2}\),
  \(\displaystyle {0.4}\),
  \(\displaystyle {0.6}\),
  \(\displaystyle {0.8}\),
  \(\displaystyle {1.0}\),
  \(\displaystyle {1.2}\)
}
]
\draw[draw=none,fill=darkslateblue6886129] (axis cs:-0.4,0) rectangle (axis cs:-0.133333333333333,1);

\draw[draw=none,fill=darkslateblue6886129] (axis cs:0.6,0) rectangle (axis cs:0.866666666666667,1);
\draw[draw=none,fill=darkslateblue6886129] (axis cs:1.6,0) rectangle (axis cs:1.86666666666667,0.969478908188586);
\draw[draw=none,fill=darkslateblue6886129] (axis cs:2.6,0) rectangle (axis cs:2.86666666666667,0.930265386787126);
\draw[draw=none,fill=darkslateblue6886129] (axis cs:3.6,0) rectangle (axis cs:3.86666666666667,0.857112571549714);
\draw[draw=none,fill=darkslateblue6886129] (axis cs:4.6,0) rectangle (axis cs:4.86666666666667,0.691006413169722);
\draw[draw=none,fill=seagreen46130127] (axis cs:-0.133333333333333,0) rectangle (axis cs:0.133333333333333,0.992925925925926);

\draw[draw=none,fill=seagreen46130127] (axis cs:0.866666666666667,0) rectangle (axis cs:1.13333333333333,0.972250043719496);
\draw[draw=none,fill=seagreen46130127] (axis cs:1.86666666666667,0) rectangle (axis cs:2.13333333333333,0.885812237484391);
\draw[draw=none,fill=seagreen46130127] (axis cs:2.86666666666667,0) rectangle (axis cs:3.13333333333333,0.740444363218631);
\draw[draw=none,fill=seagreen46130127] (axis cs:3.86666666666667,0) rectangle (axis cs:4.13333333333333,0.546749621498865);
\draw[draw=none,fill=seagreen46130127] (axis cs:4.86666666666667,0) rectangle (axis cs:5.13333333333333,0.17822566522961);
\draw[draw=none,fill=mediumseagreen107187110] (axis cs:0.133333333333333,0) rectangle (axis cs:0.4,0.981097512199024);

\draw[draw=none,fill=mediumseagreen107187110] (axis cs:1.13333333333333,0) rectangle (axis cs:1.4,0.93094526337491);
\draw[draw=none,fill=mediumseagreen107187110] (axis cs:2.13333333333333,0) rectangle (axis cs:2.4,0.754769089858272);
\draw[draw=none,fill=mediumseagreen107187110] (axis cs:3.13333333333333,0) rectangle (axis cs:3.4,0.504102450597797);
\draw[draw=none,fill=mediumseagreen107187110] (axis cs:4.13333333333333,0) rectangle (axis cs:4.4,0.310570453376427);
\draw[draw=none,fill=mediumseagreen107187110] (axis cs:5.13333333333333,0) rectangle (axis cs:5.4,0.0245436782156984);
\addplot [line width=0.72pt, darkslategray66, forget plot]
table {%
-0.266666666666667 nan
-0.266666666666667 nan
};
\addplot [line width=0.72pt, darkslategray66, forget plot]
table {%
0.733333333333333 nan
0.733333333333333 nan
};
\addplot [line width=0.72pt, darkslategray66, forget plot]
table {%
1.73333333333333 nan
1.73333333333333 nan
};
\addplot [line width=0.72pt, darkslategray66, forget plot]
table {%
2.73333333333333 nan
2.73333333333333 nan
};
\addplot [line width=0.72pt, darkslategray66, forget plot]
table {%
3.73333333333333 nan
3.73333333333333 nan
};
\addplot [line width=0.72pt, darkslategray66, forget plot]
table {%
4.73333333333333 nan
4.73333333333333 nan
};
\addplot [line width=0.72pt, darkslategray66, forget plot]
table {%
0 nan
0 nan
};
\addplot [line width=0.72pt, darkslategray66, forget plot]
table {%
1 nan
1 nan
};
\addplot [line width=0.72pt, darkslategray66, forget plot]
table {%
2 nan
2 nan
};
\addplot [line width=0.72pt, darkslategray66, forget plot]
table {%
3 nan
3 nan
};
\addplot [line width=0.72pt, darkslategray66, forget plot]
table {%
4 nan
4 nan
};
\addplot [line width=0.72pt, darkslategray66, forget plot]
table {%
5 nan
5 nan
};
\addplot [line width=0.72pt, darkslategray66, forget plot]
table {%
0.266666666666667 nan
0.266666666666667 nan
};
\addplot [line width=0.72pt, darkslategray66, forget plot]
table {%
1.26666666666667 nan
1.26666666666667 nan
};
\addplot [line width=0.72pt, darkslategray66, forget plot]
table {%
2.26666666666667 nan
2.26666666666667 nan
};
\addplot [line width=0.72pt, darkslategray66, forget plot]
table {%
3.26666666666667 nan
3.26666666666667 nan
};
\addplot [line width=0.72pt, darkslategray66, forget plot]
table {%
4.26666666666667 nan
4.26666666666667 nan
};
\addplot [line width=0.72pt, darkslategray66, forget plot]
table {%
5.26666666666667 nan
5.26666666666667 nan
};
\end{axis}

\end{tikzpicture}
    }
\end{minipage}
\caption{DR distribution, average data rate, and PRR vs. $h$ and $\theta$, with $G_{tx}+G_{rx} = 5$ dBi and $p=60$ s.}
\label{fig::dr_scenario}
\vskip -0.4cm
\end{figure*}
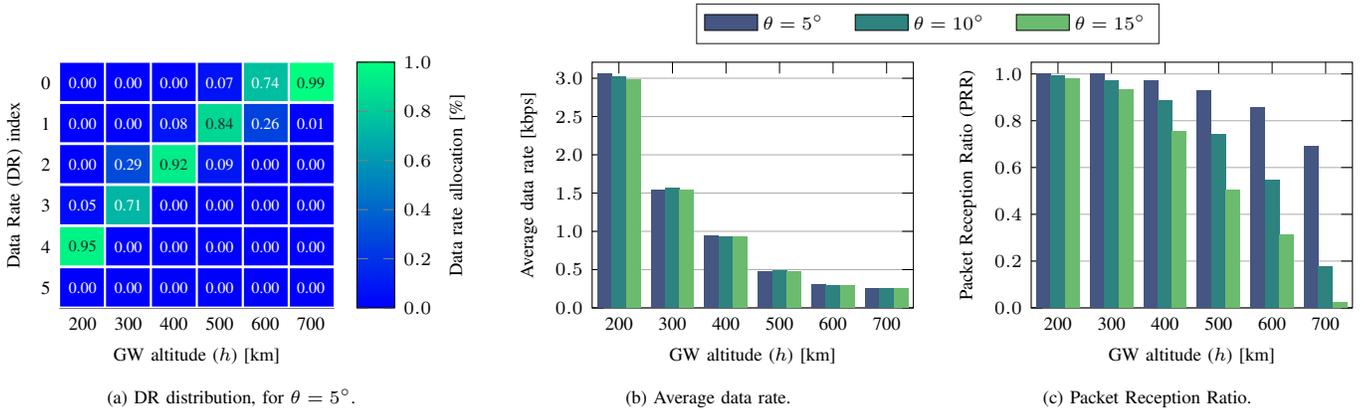

\subsection{Simulation Setup}
We run simulations using the \texttt{ns3-LoRa-NTN} module described in Sec.~\ref{sub:framework}, and simulation parameters are provided in~\cref{tab:parameters}.
We consider a rural scenario where $N$ \glspl{ed} are uniformly distributed on the Earth’s surface with a fixed spatial density $\rho_d = 0.01$~EDs/km\textsuperscript{2}.
A single \gls{leo} satellite, at altitude $h$, serves as a LoRa \gls{gw}, collecting uplink traffic from the \glspl{ed}, and forwarding it to the \gls{ns} for storage or processing.
Each \gls{ed} transmits packets of fixed size $p_s = 32$~bytes at regular intervals with a transmission periodicity of $p$.
The transmit power is fixed to $P_{tx} = 14$~dBm, and the bandwidth is $B=125$~kHz~\cite{Petajajarvi2015LoRaCoverage}.
As far as the antenna model is concerned, several studies, including~\cite{Rahman2017,Adnan2018}, assume that the antenna gain for LoRa \glspl{gw} is approximately $G_{rx}=5$~dBi, while \glspl{ed} transmit via isotropic-like antennas due to energy and space constraints, so $G_{tx}=0$~dBi. 
In contrast, the authors of~\cite{Mainsuri2020} and~\cite{s23125359} proposed an enhanced antenna model for LoRa \glspl{gw} and \glspl{ed} in the \gls{ism} band with a maximum gain of $G_{rx}=8.5$~dBi and $G_{rx}=1.9$~dBi, respectively, so we have $G_{tx}+G_{rx}\simeq10$~dBi.
Therefore, in this study we consider both models, to evaluate the impact of different antenna configurations on the network.

Simulation results are given as a function of the satellite altitude $h$, the antenna beamwidth $\theta$, and the transmission periodicity $p$, in terms of: (i) the number of covered \glspl{ed} $N$, the coverage radius $R_c$ (the ground-projected radius of the satellite footprint); (iii) the \gls{dr} distribution; and (iv) the \gls{prr} and average data rate.

 \subsection{Coverage Analysis}
 \label{sub:coverage-analysis}
 In~\cref{fig:coverage} we plot the satellite’s coverage radius ($R_c$) and the number of \glspl{ed} it can serve ($N$), both of which depend on the satellite altitude $h$ and the antenna beamwidth $\theta$.
Indeed, from \cref{eq:R_c}, $R_c$ increases with both $h$ and $\theta$. For instance, at a satellite altitude of $h = 200$~km, a beamwidth of $\theta = 5^\circ$ results in a coverage radius of approximately $10$~km, while increasing $\theta$ to $15^\circ$ expands the radius to around $25$~km. At higher altitudes, the coverage area grows further; for example, at $h = 700$~km and $\theta = 15^\circ$, $R_c\simeq90$~km.
Given that the device density $\rho_d$ is fixed,  from \cref{eq:A_c}, the number of covered EDs $N$ is directly proportional to the coverage area $A_c$.
For example, with $h = 700$~km and $\theta = 5^\circ$, the satellite covers approximately $30$~\glspl{ed}, vs. $267$~\glspl{ed} for $\theta = 15^\circ$.

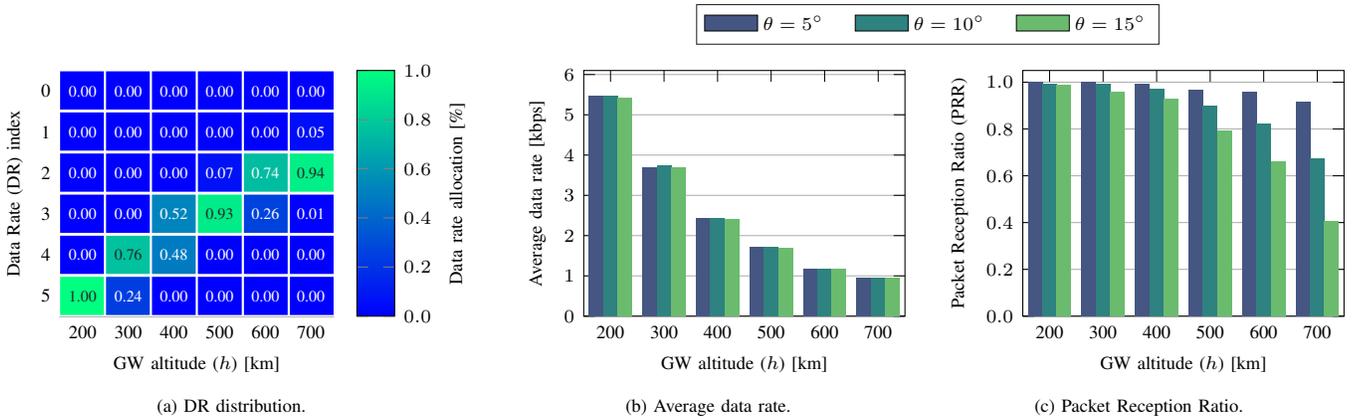
\begin{figure*}[t!]
\centering
\subfloat[DR distribution.]
{
    \label{fig:dr_distribution_2}
\begin{tikzpicture}

\definecolor{darkgray176}{RGB}{176,176,176}
\definecolor{darkslategray38}{RGB}{38,38,38}

\begin{axis}[
width = \textwidth/3.5,
height = 4.85cm,
colorbar,
colorbar style={ytick={0,0.2,0.4,0.6,0.8,1},yticklabels={
  \(\displaystyle {0.0}\),
  \(\displaystyle {0.2}\),
  \(\displaystyle {0.4}\),
  \(\displaystyle {0.6}\),
  \(\displaystyle {0.8}\),
  \(\displaystyle {1.0}\)
},ylabel={Data rate allocation [\%]}},
colormap={mymap}{[1pt]
  rgb(0pt)=(0,0,1);
  rgb(1pt)=(0,1,0.5)
},
point meta max=1,
point meta min=0,
tick pos=both,
x grid style={darkgray176},
xlabel={GW altitude ($h$) [km]},
xmin=0, xmax=6,
xtick style={color=black},
xtick={0.5,1.5,2.5,3.5,4.5,5.5},
xticklabels={200,300,400,500,600,700},
y dir=reverse,
y grid style={darkgray176},
ylabel={\acrfull{dr} index},
ymin=0, ymax=6,
ytick style={color=black},
ytick={0.5,1.5,2.5,3.5,4.5,5.5},
yticklabels={0,1,2,3,4,5}
]
\addplot graphics [includegraphics cmd=\pgfimage,xmin=0, xmax=6, ymin=6, ymax=0] {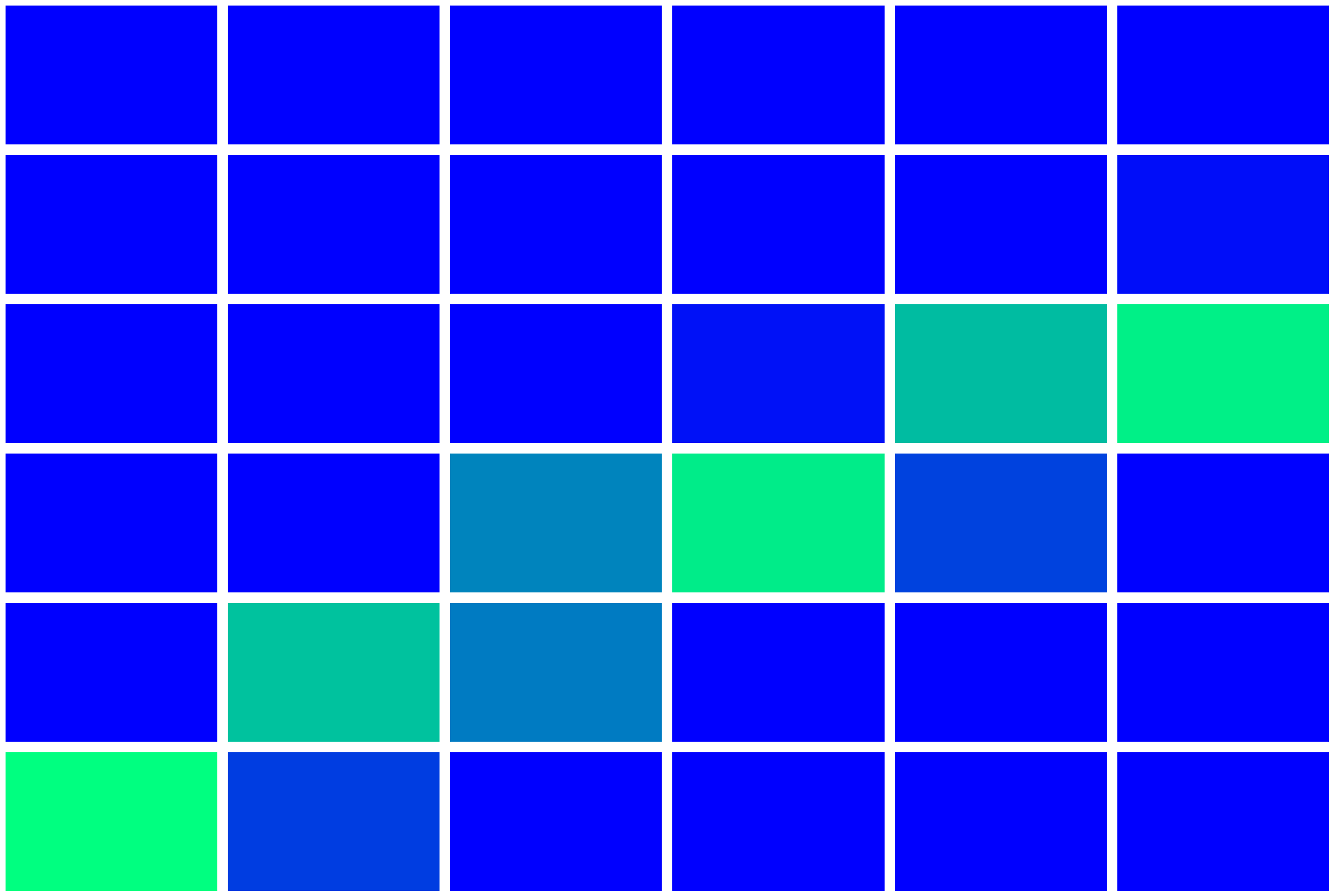};
\draw (axis cs:0.5,0.5) node[
  scale=0.6,
  text=white,
  rotate=0.0
]{0.00};
\draw (axis cs:1.5,0.5) node[
  scale=0.6,
  text=white,
  rotate=0.0
]{0.00};
\draw (axis cs:2.5,0.5) node[
  scale=0.6,
  text=white,
  rotate=0.0
]{0.00};
\draw (axis cs:3.5,0.5) node[
  scale=0.6,
  text=white,
  rotate=0.0
]{0.00};
\draw (axis cs:4.5,0.5) node[
  scale=0.6,
  text=white,
  rotate=0.0
]{0.00};
\draw (axis cs:5.5,0.5) node[
  scale=0.6,
  text=white,
  rotate=0.0
]{0.00};
\draw (axis cs:0.5,1.5) node[
  scale=0.6,
  text=white,
  rotate=0.0
]{0.00};
\draw (axis cs:1.5,1.5) node[
  scale=0.6,
  text=white,
  rotate=0.0
]{0.00};
\draw (axis cs:2.5,1.5) node[
  scale=0.6,
  text=white,
  rotate=0.0
]{0.00};
\draw (axis cs:3.5,1.5) node[
  scale=0.6,
  text=white,
  rotate=0.0
]{0.00};
\draw (axis cs:4.5,1.5) node[
  scale=0.6,
  text=white,
  rotate=0.0
]{0.00};
\draw (axis cs:5.5,1.5) node[
  scale=0.6,
  text=white,
  rotate=0.0
]{0.05};
\draw (axis cs:0.5,2.5) node[
  scale=0.6,
  text=white,
  rotate=0.0
]{0.00};
\draw (axis cs:1.5,2.5) node[
  scale=0.6,
  text=white,
  rotate=0.0
]{0.00};
\draw (axis cs:2.5,2.5) node[
  scale=0.6,
  text=white,
  rotate=0.0
]{0.00};
\draw (axis cs:3.5,2.5) node[
  scale=0.6,
  text=white,
  rotate=0.0
]{0.07};
\draw (axis cs:4.5,2.5) node[
  scale=0.6,
  text=white,
  rotate=0.0
]{0.74};
\draw (axis cs:5.5,2.5) node[
  scale=0.6,
  text=darkslategray38,
  rotate=0.0
]{0.94};
\draw (axis cs:0.5,3.5) node[
  scale=0.6,
  text=white,
  rotate=0.0
]{0.00};
\draw (axis cs:1.5,3.5) node[
  scale=0.6,
  text=white,
  rotate=0.0
]{0.00};
\draw (axis cs:2.5,3.5) node[
  scale=0.6,
  text=white,
  rotate=0.0
]{0.52};
\draw (axis cs:3.5,3.5) node[
  scale=0.6,
  text=darkslategray38,
  rotate=0.0
]{0.93};
\draw (axis cs:4.5,3.5) node[
  scale=0.6,
  text=white,
  rotate=0.0
]{0.26};
\draw (axis cs:5.5,3.5) node[
  scale=0.6,
  text=white,
  rotate=0.0
]{0.01};
\draw (axis cs:0.5,4.5) node[
  scale=0.6,
  text=white,
  rotate=0.0
]{0.00};
\draw (axis cs:1.5,4.5) node[
  scale=0.6,
  text=darkslategray38,
  rotate=0.0
]{0.76};
\draw (axis cs:2.5,4.5) node[
  scale=0.6,
  text=white,
  rotate=0.0
]{0.48};
\draw (axis cs:3.5,4.5) node[
  scale=0.6,
  text=white,
  rotate=0.0
]{0.00};
\draw (axis cs:4.5,4.5) node[
  scale=0.6,
  text=white,
  rotate=0.0
]{0.00};
\draw (axis cs:5.5,4.5) node[
  scale=0.6,
  text=white,
  rotate=0.0
]{0.00};
\draw (axis cs:0.5,5.5) node[
  scale=0.6,
  text=darkslategray38,
  rotate=0.0
]{1.00};
\draw (axis cs:1.5,5.5) node[
  scale=0.6,
  text=white,
  rotate=0.0
]{0.24};
\draw (axis cs:2.5,5.5) node[
  scale=0.6,
  text=white,
  rotate=0.0
]{0.00};
\draw (axis cs:3.5,5.5) node[
  scale=0.6,
  text=white,
  rotate=0.0
]{0.00};
\draw (axis cs:4.5,5.5) node[
  scale=0.6,
  text=white,
  rotate=0.0
]{0.00};
\draw (axis cs:5.5,5.5) node[
  scale=0.6,
  text=white,
  rotate=0.0
]{0.00};
\end{axis}

\end{tikzpicture}
}
\begin{minipage}[b]{0.64\textwidth}
    \centering
%
%

\definecolor{darkslateblue6886129}{RGB}{68,86,129}
\definecolor{mediumseagreen107187110}{RGB}{107,187,110}
\definecolor{seagreen46130127}{RGB}{46,130,127}

\begin{tikzpicture}
\pgfplotsset{every tick label/.append style={font=\scriptsize}}

\pgfplotsset{compat=1.11,
	/pgfplots/ybar legend/.style={
		/pgfplots/legend image code/.code={%
			\draw[##1,/tikz/.cd,yshift=-0.25em]
			(0cm,0cm) rectangle (20pt,0.6em);},
	},
}

\begin{axis}[%
width=0,
height=0,
at={(0,0)},
scale only axis,
xmin=0,
xmax=0,
xtick={},
ymin=0,
ymax=0,
ytick={},
axis background/.style={fill=white},
legend style={legend cell align=left,
              align=center,
              draw=white!15!black,
              at={(0.5, 1.3)},
              anchor=center,
              /tikz/every even column/.append style={column sep=1em}},
legend columns=3,
]
\addplot[ybar,ybar legend,draw=black,fill=darkslateblue6886129,line width=0.08pt]
table[row sep=crcr]{%
	0	0\\
};
\addlegendentry{$\theta=5^\circ$}

\addplot[ybar,ybar legend,draw=black,fill=seagreen46130127,line width=0.08pt]
table[row sep=crcr]{%
	0	0\\
};
\addlegendentry{$\theta=10^\circ$}

\addplot[ybar,ybar legend,draw=black,fill=mediumseagreen107187110,line width=0.08pt]
table[row sep=crcr]{%
	0	0\\
};
\addlegendentry{$\theta=15^\circ$}

\end{axis}
\end{tikzpicture}%
    \vskip 0.2 cm
    \subfloat[Average data rate.]
    {
        \label{fig:dr_avg_2}
\begin{tikzpicture}

\definecolor{darkgray176}{RGB}{176,176,176}
\definecolor{darkslateblue6886129}{RGB}{68,86,129}
\definecolor{darkslategray66}{RGB}{66,66,66}
\definecolor{lightgray204}{RGB}{204,204,204}
\definecolor{mediumseagreen107187110}{RGB}{107,187,110}
\definecolor{seagreen46130127}{RGB}{46,130,127}

\pgfplotsset{compat=1.11,
	/pgfplots/ybar legend/.style={
		/pgfplots/legend image code/.code={%
			\draw[##1,/tikz/.cd,yshift=-0.25em]
			(0cm,0cm) rectangle (20pt,0.6em);},
	},
}

\begin{axis}[
width = \textwidth/2,
height = 4.85cm,
legend cell align={left},
legend columns=3,
legend style={
  fill opacity=0.8,
  draw opacity=1,
  text opacity=1,
  at={(0.5,1.18)},
  anchor=north,
  draw=lightgray204,
  /tikz/every even column/.append style={column sep=0.5em}
},
tick pos=both,
unbounded coords=jump,
x grid style={darkgray176},
xlabel={GW altitude ($h$) [km]},
xmin=-0.5, xmax=5.5,
xtick style={color=black},
xtick={0,1,2,3,4,5},
xticklabels={200,300,400,500,600,700},
y grid style={darkgray176},
ymajorgrids,
ylabel={Average data rate [kbps]},
ymin=0, ymax=6.1,
ytick style={color=black},
ytick={0,1,2,3,4,5,6},
]
\draw[draw=none,fill=darkslateblue6886129] (axis cs:-0.4,0) rectangle (axis cs:-0.133333333333333,5.47);

\draw[draw=none,fill=darkslateblue6886129] (axis cs:0.6,0) rectangle (axis cs:0.866666666666667,3.6878);
\draw[draw=none,fill=darkslateblue6886129] (axis cs:1.6,0) rectangle (axis cs:1.86666666666667,2.41975);
\draw[draw=none,fill=darkslateblue6886129] (axis cs:2.6,0) rectangle (axis cs:2.86666666666667,1.70655357142857);
\draw[draw=none,fill=darkslateblue6886129] (axis cs:3.6,0) rectangle (axis cs:3.86666666666667,1.17940476190476);
\draw[draw=none,fill=darkslateblue6886129] (axis cs:4.6,0) rectangle (axis cs:4.86666666666667,0.952655172413793);
\draw[draw=none,fill=seagreen46130127] (axis cs:-0.133333333333333,0) rectangle (axis cs:0.133333333333333,5.47);

\draw[draw=none,fill=seagreen46130127] (axis cs:0.866666666666667,0) rectangle (axis cs:1.13333333333333,3.74582142857143);
\draw[draw=none,fill=seagreen46130127] (axis cs:1.86666666666667,0) rectangle (axis cs:2.13333333333333,2.42538157894737);
\draw[draw=none,fill=seagreen46130127] (axis cs:2.86666666666667,0) rectangle (axis cs:3.13333333333333,1.70674583333333);
\draw[draw=none,fill=seagreen46130127] (axis cs:3.86666666666667,0) rectangle (axis cs:4.13333333333333,1.17044186046512);
\draw[draw=none,fill=seagreen46130127] (axis cs:4.86666666666667,0) rectangle (axis cs:5.13333333333333,0.95167094017094);
\draw[draw=none,fill=mediumseagreen107187110] (axis cs:0.133333333333333,0) rectangle (axis cs:0.4,5.40830952380952);

\draw[draw=none,fill=mediumseagreen107187110] (axis cs:1.13333333333333,0) rectangle (axis cs:1.4,3.68997448979592);
\draw[draw=none,fill=mediumseagreen107187110] (axis cs:2.13333333333333,0) rectangle (axis cs:2.4,2.39906896551724);
\draw[draw=none,fill=mediumseagreen107187110] (axis cs:3.13333333333333,0) rectangle (axis cs:3.4,1.68664338235294);
\draw[draw=none,fill=mediumseagreen107187110] (axis cs:4.13333333333333,0) rectangle (axis cs:4.4,1.16091709183673);
\draw[draw=none,fill=mediumseagreen107187110] (axis cs:5.13333333333333,0) rectangle (axis cs:5.4,0.947568609022556);
\addplot [line width=0.72pt, darkslategray66, forget plot]
table {%
-0.266666666666667 nan
-0.266666666666667 nan
};
\addplot [line width=0.72pt, darkslategray66, forget plot]
table {%
0.733333333333333 nan
0.733333333333333 nan
};
\addplot [line width=0.72pt, darkslategray66, forget plot]
table {%
1.73333333333333 nan
1.73333333333333 nan
};
\addplot [line width=0.72pt, darkslategray66, forget plot]
table {%
2.73333333333333 nan
2.73333333333333 nan
};
\addplot [line width=0.72pt, darkslategray66, forget plot]
table {%
3.73333333333333 nan
3.73333333333333 nan
};
\addplot [line width=0.72pt, darkslategray66, forget plot]
table {%
4.73333333333333 nan
4.73333333333333 nan
};
\addplot [line width=0.72pt, darkslategray66, forget plot]
table {%
0 nan
0 nan
};
\addplot [line width=0.72pt, darkslategray66, forget plot]
table {%
1 nan
1 nan
};
\addplot [line width=0.72pt, darkslategray66, forget plot]
table {%
2 nan
2 nan
};
\addplot [line width=0.72pt, darkslategray66, forget plot]
table {%
3 nan
3 nan
};
\addplot [line width=0.72pt, darkslategray66, forget plot]
table {%
4 nan
4 nan
};
\addplot [line width=0.72pt, darkslategray66, forget plot]
table {%
5 nan
5 nan
};
\addplot [line width=0.72pt, darkslategray66, forget plot]
table {%
0.266666666666667 nan
0.266666666666667 nan
};
\addplot [line width=0.72pt, darkslategray66, forget plot]
table {%
1.26666666666667 nan
1.26666666666667 nan
};
\addplot [line width=0.72pt, darkslategray66, forget plot]
table {%
2.26666666666667 nan
2.26666666666667 nan
};
\addplot [line width=0.72pt, darkslategray66, forget plot]
table {%
3.26666666666667 nan
3.26666666666667 nan
};
\addplot [line width=0.72pt, darkslategray66, forget plot]
table {%
4.26666666666667 nan
4.26666666666667 nan
};
\addplot [line width=0.72pt, darkslategray66, forget plot]
table {%
5.26666666666667 nan
5.26666666666667 nan
};
\end{axis}

\end{tikzpicture}
    }
    \subfloat[Packet Reception Ratio.]
    {
        \label{fig:prr_2}
\begin{tikzpicture}

\definecolor{darkgray176}{RGB}{176,176,176}
\definecolor{darkslateblue6886129}{RGB}{68,86,129}
\definecolor{darkslategray66}{RGB}{66,66,66}
\definecolor{lightgray204}{RGB}{204,204,204}
\definecolor{mediumseagreen107187110}{RGB}{107,187,110}
\definecolor{seagreen46130127}{RGB}{46,130,127}

\pgfplotsset{compat=1.11,
	/pgfplots/ybar legend/.style={
		/pgfplots/legend image code/.code={%
			\draw[##1,/tikz/.cd,yshift=-0.25em]
			(0cm,0cm) rectangle (20pt,0.6em);},
	},
}

\begin{axis}[
width = \textwidth/2,
height = 4.85cm,
legend cell align={left},
legend columns=3,
legend style={
  fill opacity=0.8,
  draw opacity=1,
  text opacity=1,
  at={(0.5,1.18)},
  anchor=north,
  draw=lightgray204,
  /tikz/every even column/.append style={column sep=0.5em}
},
tick pos=both,
unbounded coords=jump,
x grid style={darkgray176},
xlabel={GW altitude ($h$) [km]},
xmin=-0.5, xmax=5.5,
xtick style={color=black},
xtick={0,1,2,3,4,5},
xticklabels={200,300,400,500,600,700},
y grid style={darkgray176},
ylabel={Packet Reception Ratio (PRR)},
ymajorgrids,
ymin=0, ymax=1.05,
ytick style={color=black},
ytick={0,0.2,0.4,0.6,0.8,1,1.2},
yticklabels={
  \(\displaystyle {0.0}\),
  \(\displaystyle {0.2}\),
  \(\displaystyle {0.4}\),
  \(\displaystyle {0.6}\),
  \(\displaystyle {0.8}\),
  \(\displaystyle {1.0}\),
  \(\displaystyle {1.2}\)
}
]
\draw[draw=none,fill=darkslateblue6886129] (axis cs:-0.4,0) rectangle (axis cs:-0.133333333333333,1);

\draw[draw=none,fill=darkslateblue6886129] (axis cs:0.6,0) rectangle (axis cs:0.866666666666667,1);
\draw[draw=none,fill=darkslateblue6886129] (axis cs:1.6,0) rectangle (axis cs:1.86666666666667,0.99237037037037);
\draw[draw=none,fill=darkslateblue6886129] (axis cs:2.6,0) rectangle (axis cs:2.86666666666667,0.967399346077659);
\draw[draw=none,fill=darkslateblue6886129] (axis cs:3.6,0) rectangle (axis cs:3.86666666666667,0.957362336598057);
\draw[draw=none,fill=darkslateblue6886129] (axis cs:4.6,0) rectangle (axis cs:4.86666666666667,0.915448808802686);
\draw[draw=none,fill=seagreen46130127] (axis cs:-0.133333333333333,0) rectangle (axis cs:0.133333333333333,0.992925925925926);

\draw[draw=none,fill=seagreen46130127] (axis cs:0.866666666666667,0) rectangle (axis cs:1.13333333333333,0.991287896853786);
\draw[draw=none,fill=seagreen46130127] (axis cs:1.86666666666667,0) rectangle (axis cs:2.13333333333333,0.972181842209594);
\draw[draw=none,fill=seagreen46130127] (axis cs:2.86666666666667,0) rectangle (axis cs:3.13333333333333,0.899350993008551);
\draw[draw=none,fill=seagreen46130127] (axis cs:3.86666666666667,0) rectangle (axis cs:4.13333333333333,0.824300211128516);
\draw[draw=none,fill=seagreen46130127] (axis cs:4.86666666666667,0) rectangle (axis cs:5.13333333333333,0.674791263284241);
\draw[draw=none,fill=mediumseagreen107187110] (axis cs:0.133333333333333,0) rectangle (axis cs:0.4,0.989216023278681);

\draw[draw=none,fill=mediumseagreen107187110] (axis cs:1.13333333333333,0) rectangle (axis cs:1.4,0.960073923184468);
\draw[draw=none,fill=mediumseagreen107187110] (axis cs:2.13333333333333,0) rectangle (axis cs:2.4,0.928597716645553);
\draw[draw=none,fill=mediumseagreen107187110] (axis cs:3.13333333333333,0) rectangle (axis cs:3.4,0.793023642321517);
\draw[draw=none,fill=mediumseagreen107187110] (axis cs:4.13333333333333,0) rectangle (axis cs:4.4,0.659064842343466);
\draw[draw=none,fill=mediumseagreen107187110] (axis cs:5.13333333333333,0) rectangle (axis cs:5.4,0.405405178531417);
\addplot [line width=0.72pt, darkslategray66, forget plot]
table {%
-0.266666666666667 nan
-0.266666666666667 nan
};
\addplot [line width=0.72pt, darkslategray66, forget plot]
table {%
0.733333333333333 nan
0.733333333333333 nan
};
\addplot [line width=0.72pt, darkslategray66, forget plot]
table {%
1.73333333333333 nan
1.73333333333333 nan
};
\addplot [line width=0.72pt, darkslategray66, forget plot]
table {%
2.73333333333333 nan
2.73333333333333 nan
};
\addplot [line width=0.72pt, darkslategray66, forget plot]
table {%
3.73333333333333 nan
3.73333333333333 nan
};
\addplot [line width=0.72pt, darkslategray66, forget plot]
table {%
4.73333333333333 nan
4.73333333333333 nan
};
\addplot [line width=0.72pt, darkslategray66, forget plot]
table {%
0 nan
0 nan
};
\addplot [line width=0.72pt, darkslategray66, forget plot]
table {%
1 nan
1 nan
};
\addplot [line width=0.72pt, darkslategray66, forget plot]
table {%
2 nan
2 nan
};
\addplot [line width=0.72pt, darkslategray66, forget plot]
table {%
3 nan
3 nan
};
\addplot [line width=0.72pt, darkslategray66, forget plot]
table {%
4 nan
4 nan
};
\addplot [line width=0.72pt, darkslategray66, forget plot]
table {%
5 nan
5 nan
};
\addplot [line width=0.72pt, darkslategray66, forget plot]
table {%
0.266666666666667 nan
0.266666666666667 nan
};
\addplot [line width=0.72pt, darkslategray66, forget plot]
table {%
1.26666666666667 nan
1.26666666666667 nan
};
\addplot [line width=0.72pt, darkslategray66, forget plot]
table {%
2.26666666666667 nan
2.26666666666667 nan
};
\addplot [line width=0.72pt, darkslategray66, forget plot]
table {%
3.26666666666667 nan
3.26666666666667 nan
};
\addplot [line width=0.72pt, darkslategray66, forget plot]
table {%
4.26666666666667 nan
4.26666666666667 nan
};
\addplot [line width=0.72pt, darkslategray66, forget plot]
table {%
5.26666666666667 nan
5.26666666666667 nan
};
\end{axis}

\end{tikzpicture}
    }
\end{minipage}
\caption{DR distribution, average data rate, and PRR vs. $h$ and $\theta$, with $G_{tx}+G_{rx} = 10$ dBi and $p=60$ s.}
\label{fig::dr_scenario}
\vskip -0.4cm
\end{figure*}

\subsection{Limited Antenna Gain}
\label{sec:limited_gain}

In this analysis, the total antenna gain is set to $G_{tx}+G_{rx}=5$~dBi, which reflects the typical gain used in standard LoRa gateway configurations~\cite{Rahman2017,Adnan2018}.
\cref{fig:dr_distribution_1} illustrates the DR distribution (which directly depends on the corresponding \gls{sf}, as reported in~\cref{tab:lora_sf_parameters}), for a beamwidth of $\theta = 5^\circ$, as a function of the satellite altitude $h$.
On one side, increasing the altitude expands the satellite's coverage area. 
On the other side, the resulting longer communication distance introduces more path loss, which degrades the signal quality. As such, \glspl{ed} are forced to use a higher SF to operate at lower sensitivity and increase the coverage range, so the resulting DR index decreases.
For instance, at $h = 200$~km, approximately 95\% of the \glspl{ed} operate with $\text{DR} = 4$ ($\text{SF} = 8$), and the corresponding average data rate is about 3~kbps, as shown in~\cref{fig:dr_avg_1}.
For $h>500$~km, most EDs select $\text{DR} = 0$, and use the same $\text{SF} = 12$, which significantly increases the probability of collision. The corresponding average data rate is only 300 bps. At $h = 700$~km, nearly all \glspl{ed} operate with $\text{DR} = 0$. 

Moreover, the impact of the beamwidth $\theta$ is not negligible.
As $\theta$ increases, the number of covered EDs $N$ also increases, so the \gls{prr} decreases, as illustrated in~\cref{fig:prr_1}. For example, see that the \gls{prr} drops to approximately 0.7 for $\theta = 5^\circ$ (where $N=30$), and further decreases to just 0.02 for $\theta = 15^\circ$ (where $N=267$). This degradation occurs because the probability of packet collision increases with $N$, especially when many EDs use the same SF.
Naturally, this behavior also depends on $h$. 
For instance, at $h = 400$~km and with $\theta = 5^\circ$, the satellite footprint has a radius of approximately 17~km, covering only about $N=10$ \glspl{ed}, so the PRR is 0.97.
In contrast, with $h = 700$~km, the radius increases to approximately 30~km and $N=30$ EDs, so the PRR decreases to 0.7.

\subsection{Enhanced Antenna Gain}
\label{sec:enhanced_gain}

Now, we evaluate the network performance assuming a total antenna gain of $G_{tx}+G_{rx}=10$~dBi, thereby using the enhanced configuration proposed in~\cite{Mainsuri2020} and~\cite{s23125359} .
As shown in~\cref{fig:dr_distribution_2}, the additional antenna gain significantly improves the link quality compared to the scenario in \cref{sec:limited_gain}.
At an altitude of $h = 200$~km, all \glspl{ed} select $\text{DR} = 5$ ($\text{SF} = 7$), corresponding to the highest data rate supported by \gls{lora} (approximately $5.47$~kbps), as illustrated in~\cref{fig:dr_avg_2}. 
Even at $h = 300$~km, $24$\% of the EDs still maintain $\text{DR} = 5$, while the rest switch to a slightly lower DR index (i.e., higher SF) to satisfy the sensitivity requirements of the gateway.
From $h = 400$~km, $\text{DR} = 5$ is no longer feasible due to more severe path loss.
A notable improvement with respect to the previous scenario is observed at $h = 500$~km: in these conditions, most \glspl{ed} operate with $\text{DR} = 3$ ($\text{SF} = 9$) when $G_{tx}+G_{rx}=10$~dBi, in contrast to $\text{DR} = 1$ when $G_{tx}+G_{rx}=5$~dBi.
The resulting data rate increases by more than two times, from around 1 kbps to 2.5 kbps.
The PRR also improves to around $0.8$, vs. $0.5$.
At $h = 600$~km, $\text{DR} = 2$ ($\text{SF} = 10$) becomes the most common selection, so the average data rate decreases to only 1.8 kbps.
Nevertheless, the \gls{prr} remains around $0.91$ for $\theta = 5^\circ$, and approximately $0.66$ when $\theta = 15^\circ$.

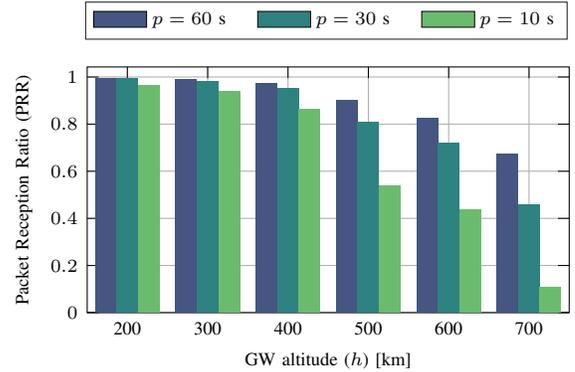
\begin{figure}[t!]
\centering
\makebox[\linewidth][l]{\hspace{0.2cm}
%
%

\definecolor{brown1794649}{RGB}{179,46,49}
\definecolor{darkgray176}{RGB}{176,176,176}
\definecolor{darkslategray66}{RGB}{66,66,66}
\definecolor{lightgray204}{RGB}{204,204,204}
\definecolor{lightpink240191172}{RGB}{240,191,172}
\definecolor{salmon22811995}{RGB}{228,119,95}
\definecolor{darkgray176}{RGB}{176,176,176}
\definecolor{darkslateblue6886129}{RGB}{68,86,129}
\definecolor{darkslategray66}{RGB}{66,66,66}
\definecolor{lightgray204}{RGB}{204,204,204}
\definecolor{mediumseagreen107187110}{RGB}{107,187,110}
\definecolor{seagreen46130127}{RGB}{46,130,127}

\begin{tikzpicture}
\pgfplotsset{every tick label/.append style={font=\scriptsize}}

\pgfplotsset{compat=1.11,
	/pgfplots/ybar legend/.style={
		/pgfplots/legend image code/.code={%
			\draw[##1,/tikz/.cd,yshift=-0.25em]
			(0cm,0cm) rectangle (20pt,0.6em);},
	},
}

\begin{axis}[%
width=0,
height=0,
at={(0,0)},
scale only axis,
xmin=0,
xmax=0,
xtick={},
ymin=0,
ymax=0,
ytick={},
axis background/.style={fill=white},
legend style={legend cell align=left,
              align=center,
              draw=white!15!black,
              at={(0.5, 1.3)},
              anchor=center,
              /tikz/every even column/.append style={column sep=1em}},
legend columns=3,
]
\addplot[ybar,ybar legend,draw=black,fill=darkslateblue6886129,line width=0.08pt]
table[row sep=crcr]{%
	0	0\\
};
\addlegendentry{$p=60$~s}

\addplot[ybar,ybar legend,draw=black,fill=seagreen46130127,line width=0.08pt]
table[row sep=crcr]{%
	0	0\\
};
\addlegendentry{$p=30$~s}

\addplot[ybar,ybar legend,draw=black,fill=mediumseagreen107187110,line width=0.08pt]
table[row sep=crcr]{%
	0	0\\
};
\addlegendentry{$p=10$~s}

\end{axis}
\end{tikzpicture}%
}
\vskip 0.3cm
\begin{tikzpicture}

\definecolor{brown1794649}{RGB}{179,46,49}
\definecolor{darkgray176}{RGB}{176,176,176}
\definecolor{darkslategray66}{RGB}{66,66,66}
\definecolor{lightgray204}{RGB}{204,204,204}
\definecolor{lightpink240191172}{RGB}{240,191,172}
\definecolor{salmon22811995}{RGB}{228,119,95}
\definecolor{darkgray176}{RGB}{176,176,176}
\definecolor{darkslateblue6886129}{RGB}{68,86,129}
\definecolor{darkslategray66}{RGB}{66,66,66}
\definecolor{lightgray204}{RGB}{204,204,204}
\definecolor{mediumseagreen107187110}{RGB}{107,187,110}
\definecolor{seagreen46130127}{RGB}{46,130,127}

\begin{axis}[
width = 0.9\columnwidth,
height = 4.85cm,
legend cell align={left},
legend style={fill opacity=0.8, draw opacity=1, text opacity=1, draw=lightgray204},
tick pos=both,
unbounded coords=jump,
x grid style={darkgray176},
xlabel={GW altitude ($h$) [km]},
xmajorgrids,
xmin=-0.5, xmax=5.5,
xtick style={color=black},
xtick={0,1,2,3,4,5},
xticklabels={200,300,400,500,600,700},
y grid style={darkgray176},
ylabel={Packet Reception Ratio (PRR)},
ymajorgrids,
ymin=0, ymax=1.04257222222222,
ytick style={color=black},
ytick={0,0.2,0.4,0.6,0.8,1,1.2},
]
\draw[draw=none,fill=darkslateblue6886129] (axis cs:-0.4,0) rectangle (axis cs:-0.133333333333333,0.992925925925926);

\draw[draw=none,fill=darkslateblue6886129] (axis cs:0.6,0) rectangle (axis cs:0.866666666666667,0.991287896853786);
\draw[draw=none,fill=darkslateblue6886129] (axis cs:1.6,0) rectangle (axis cs:1.86666666666667,0.972181842209594);
\draw[draw=none,fill=darkslateblue6886129] (axis cs:2.6,0) rectangle (axis cs:2.86666666666667,0.899350993008551);
\draw[draw=none,fill=darkslateblue6886129] (axis cs:3.6,0) rectangle (axis cs:3.86666666666667,0.824300211128516);
\draw[draw=none,fill=darkslateblue6886129] (axis cs:4.6,0) rectangle (axis cs:4.86666666666667,0.674791263284241);
\draw[draw=none,fill=seagreen46130127] (axis cs:-0.133333333333333,0) rectangle (axis cs:0.133333333333333,0.9925);

\draw[draw=none,fill=seagreen46130127] (axis cs:0.866666666666667,0) rectangle (axis cs:1.13333333333333,0.979888827921846);
\draw[draw=none,fill=seagreen46130127] (axis cs:1.86666666666667,0) rectangle (axis cs:2.13333333333333,0.951388062005248);
\draw[draw=none,fill=seagreen46130127] (axis cs:2.86666666666667,0) rectangle (axis cs:3.13333333333333,0.807845502030017);
\draw[draw=none,fill=seagreen46130127] (axis cs:3.86666666666667,0) rectangle (axis cs:4.13333333333333,0.721456806182866);
\draw[draw=none,fill=seagreen46130127] (axis cs:4.86666666666667,0) rectangle (axis cs:5.13333333333333,0.459157612746517);
\draw[draw=none,fill=mediumseagreen107187110] (axis cs:0.133333333333333,0) rectangle (axis cs:0.4,0.965509259259259);

\draw[draw=none,fill=mediumseagreen107187110] (axis cs:1.13333333333333,0) rectangle (axis cs:1.4,0.937340254623837);
\draw[draw=none,fill=mediumseagreen107187110] (axis cs:2.13333333333333,0) rectangle (axis cs:2.4,0.862734674498831);
\draw[draw=none,fill=mediumseagreen107187110] (axis cs:3.13333333333333,0) rectangle (axis cs:3.4,0.536869757688723);
\draw[draw=none,fill=mediumseagreen107187110] (axis cs:4.13333333333333,0) rectangle (axis cs:4.4,0.435584418387425);
\draw[draw=none,fill=mediumseagreen107187110] (axis cs:5.13333333333333,0) rectangle (axis cs:5.4,0.106289543539763);
\addplot [line width=0.72pt, darkslategray66, forget plot]
table {%
-0.266666666666667 nan
-0.266666666666667 nan
};
\addplot [line width=0.72pt, darkslategray66, forget plot]
table {%
0.733333333333333 nan
0.733333333333333 nan
};
\addplot [line width=0.72pt, darkslategray66, forget plot]
table {%
1.73333333333333 nan
1.73333333333333 nan
};
\addplot [line width=0.72pt, darkslategray66, forget plot]
table {%
2.73333333333333 nan
2.73333333333333 nan
};
\addplot [line width=0.72pt, darkslategray66, forget plot]
table {%
3.73333333333333 nan
3.73333333333333 nan
};
\addplot [line width=0.72pt, darkslategray66, forget plot]
table {%
4.73333333333333 nan
4.73333333333333 nan
};
\addplot [line width=0.72pt, darkslategray66, forget plot]
table {%
0 nan
0 nan
};
\addplot [line width=0.72pt, darkslategray66, forget plot]
table {%
1 nan
1 nan
};
\addplot [line width=0.72pt, darkslategray66, forget plot]
table {%
2 nan
2 nan
};
\addplot [line width=0.72pt, darkslategray66, forget plot]
table {%
3 nan
3 nan
};
\addplot [line width=0.72pt, darkslategray66, forget plot]
table {%
4 nan
4 nan
};
\addplot [line width=0.72pt, darkslategray66, forget plot]
table {%
5 nan
5 nan
};
\addplot [line width=0.72pt, darkslategray66, forget plot]
table {%
0.266666666666667 nan
0.266666666666667 nan
};
\addplot [line width=0.72pt, darkslategray66, forget plot]
table {%
1.26666666666667 nan
1.26666666666667 nan
};
\addplot [line width=0.72pt, darkslategray66, forget plot]
table {%
2.26666666666667 nan
2.26666666666667 nan
};
\addplot [line width=0.72pt, darkslategray66, forget plot]
table {%
3.26666666666667 nan
3.26666666666667 nan
};
\addplot [line width=0.72pt, darkslategray66, forget plot]
table {%
4.26666666666667 nan
4.26666666666667 nan
};
\addplot [line width=0.72pt, darkslategray66, forget plot]
table {%
5.26666666666667 nan
5.26666666666667 nan
};
\end{axis}

\end{tikzpicture}
\caption{PRR vs. $p$ and~$h$, with $G_{tx}+G_{rx} = 10$ dBi and $\theta=10^\circ$.}
\label{fig:app}
\vskip -0.6cm
\end{figure}

Finally, in~\cref{fig:app} we investigate the capacity of LoRa as a function of the traffic. 
 Specifically, we change the transmission periodicity $p$ from 60 to 10 s so that, given a fixed payload of $p_s=32$ bytes, the resulting source rate increases accordingly. 
This is consistent with the analysis in ~\cite{8863372}, where the authors change the source rate from 1 packet every second to 1 packet every 100 seconds.
As expected, the PRR decreases as the traffic increases. However, while the degradation is as small as 2.5\% at $h=200$ km, it is up to 85\% at $h=700$ km.
The impact of $h$ is twofold. 
First, as $h$ increases, the number of EDs $N$ within the satellite footprint increases, so the network is more congested due to more frequent transmission attempts.
Second, as $h$ increases, the round-trip delay and the effective time on air of data packets increase.
Given that LoRaWAN uses ALOHA for channel access, 
the probability of packet collisions also increases during that time, thereby reducing the PRR.
For example, for $p=60$ s, the ED source rate is 4 bps. 
When $h = 700$~km and $\theta = 10^\circ$, we have $N\simeq 120$~\glspl{ed}, so the total source rate is around 480 bps.
For $p=10$ s, the total source rate increases to around 3 kbps.
In turn, the average data rate of LoRa is 950 bps $\ll3$ kbps (see~\cref{fig:dr_avg_2}), that now the network cannot sustain.

These results confirm that \gls{lora} networks can effectively operate in \gls{ntn} environments, particularly when enhanced antenna configurations are employed. The additional gain extends the feasible operating altitude of the LEO satellite, and improves network robustness, even in scenarios where many EDs are covered.
The performance can be further improved by leveraging the quasi-orthogonal nature of SFs. For example, EDs can spread across different SFs to reduce the impact of collisions, regardless of the value of the sensitivity, especially in the limited antenna gain scenario, and for $h<400$ km, when all EDs would otherwise select the same SF.

\section{Conclusions and Future Work}
\label{sec:conclusions}

In this work we evaluated the feasibility of \gls{lora} \glspl{lpwan} integrated with \glspl{ntn}. Using an extended version of the \texttt{ns3-LoRa} module that incorporates the \gls{3gpp} \gls{ntn} channel model, we assessed the end-to-end performance of a scenario where a single \gls{leo} satellite acts as a \gls{lora} gateway for rural \gls{iot} deployments.
Our results show that both the satellite altitude and the antenna beamwidth significantly influence coverage, data rate, and PRR.
We observed that increasing the gateway altitude extends the satellite’s footprint, thereby enlarging the coverage area. However, this also leads to higher path loss, which forces end devices to select higher \glspl{sf}. As a result, the time on air increases, and so do the probability of packet collision and the network congestion.

As part of our future work, we will address current limitations by improving the SF selection strategy to reduce the collision probability in dense \gls{ntn} scenarios.
In addition, we plan to enhance our simulation framework by including multi-satellite and multi-layered \gls{ntn} configurations, along with a realistic satellite mobility.

\section*{Acknowledgments}
This work has been partially funded by ESA within the framework of the SatNex V project, Activity INVENTIVE. The views expressed herein can in no way be taken to reflect the official opinion of ESA. The authors would like to thank Dr. Maria Rita Palattella and Prof.  Giovanni Giambene for their valuable expertise and technical support throughout the preparation of this work. 
This work was also partially supported by the European Commission through the European Union’s Horizon Europe Research and Innovation Programme under the Marie Skłodowska-Curie-SE, Grant Agreement No. 101129618, UNITE.
This research was also partially supported by the European Union under the Italian National Recovery and Resilience Plan (NRRP) of NextGenerationEU, partnership on “Telecommunications of the Future” (PE0000001 - program “RESTART”).

\bibliographystyle{IEEEtran}
\bibliography{bibliography.bib}

\end{document}